\documentclass[twocolumn,prb,showpacs,amsmath
  ,superscriptaddress]{revtex4}
\usepackage{epsfig,graphicx}

\def\be{\begin{equation}}
\def\ee{\end{equation}}
\def\e#1{\label{#1}\end{equation}}
\def\bea{\begin{eqnarray}}
\def\eea{\end{eqnarray}}
\def\ea#1{\label{#1}\end{eqnarray}}
\def\bes#1{\begin{subequations}\label{#1}}
\def\ese{\end{subequations}}

\begin{document}
\title{Theory of measurement crosstalk in superconducting phase
qubits}
\author{Abraham G. Kofman}
\altaffiliation[Permanent address: ]{Department of Chemical Physics,
The Weizmann Institute of Science, Rehovot 76100, Israel}
\author{Qin Zhang}
 \affiliation{Department of Electrical Engineering, University of
California, Riverside, California 92521}
\author{John M. Martinis}
 \affiliation{Department of Physics, University of
California, Santa Barbara, California 93106}
\author{Alexander N. Korotkov}
 \affiliation{Department of Electrical Engineering, University of
California, Riverside, California 92521}
\date{\today}

\begin{abstract}

We analyze the crosstalk error mechanism in measurement of two capacitively
coupled superconducting flux-biased phase qubits. The damped oscillations of
the superconducting phase after the measurement of the first qubit may
significantly excite the second qubit, leading to its measurement error. The
first qubit, which is highly excited after the measurement, is described
classically. The second qubit is treated both classically and
quantum-mechanically. The results of the analysis are used to find the upper
limit for the coupling capacitance (thus limiting the frequency of two-qubit
operations) for a given tolerable value of the measurement error probability.
\end{abstract}


\maketitle

\section{Introduction}

\begin{figure}[t]
\includegraphics[width=4cm]{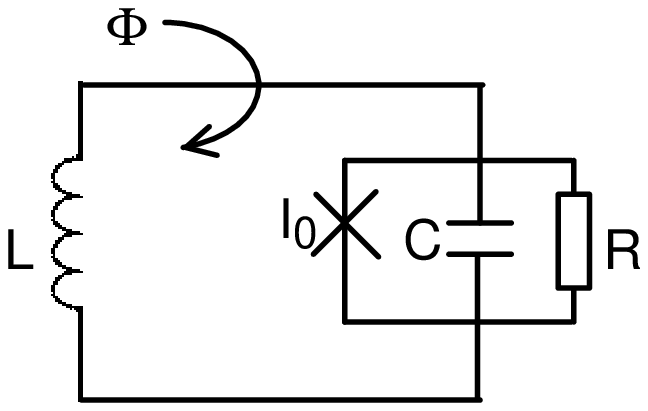}
\hspace{0.2cm}
\includegraphics[width=4cm]{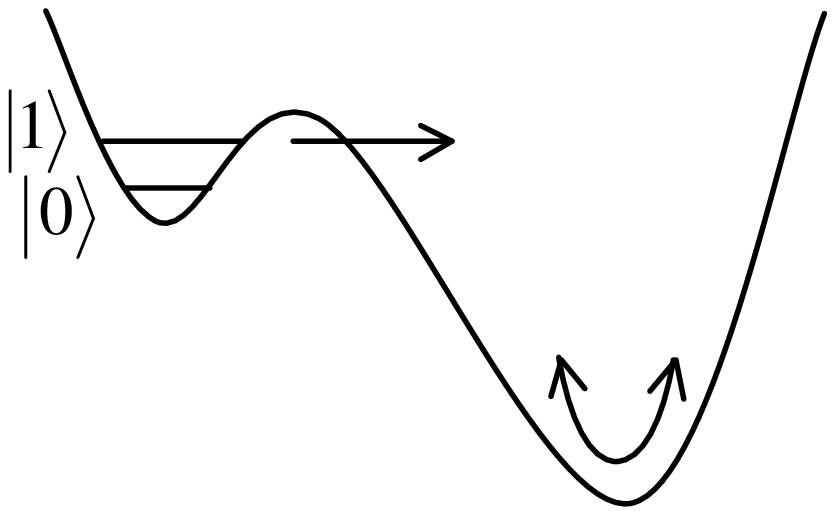}
\caption{The circuit schematic of a flux-biased phase qubit and the
corresponding potential profile (as a function of the phase difference
$\delta$ across the Josephson junction). During the measurement the state
$|1\rangle$ escapes from the ``left'' well through the barrier, which is
followed by oscillations in the ``right'' well. This dissipative evolution
leads to the two-qubit crosstalk.}
 \label{f0}\end{figure}

Superconducting Josephson-junction circuits, including phase, \cite{mar02}
flux, \cite{chi03} and charge \cite{pas03} qubits, have attracted a
significant interest as promising devices for quantum information
processing. \cite{nie00} In this paper we consider flux-biased phase qubits,
\cite{sim04,coo04,joh05} which have been introduced relatively recently and
have a clear advantage over the current-biased phase qubits. While the
schematic of a flux-biased phase qubit (Fig.\ \ref{f0}) may be very similar
to a flux qubit (in the simplest case, a superconducting loop interrupted by
one Josephson junction), an important difference is that in the phase qubit
the logic states are represented by two lowest levels in one well of the
corresponding potential profile, while for the flux qubit the levels in two
neighboring wells are used.
 An imaginary-swap quantum gate, which together with single-qubit
rotations forms a universal set of quantum gates, \cite{kem02} has been
realized with flux-biased phase qubits in Ref.\ \onlinecite{mcd05}.

Simultaneous measurement of all the qubits is an important stage in quantum
information processing, representing one of numerous challenges on the way to
a scalable quantum computer.
 Recently a scheme allowing for fast simultaneous measurement
of two coupled flux-biased phase qubits has been implemented. \cite{mcd05}
 According to this scheme, a measurement is performed by lowering the
barrier between the shallow (``left'') potential well used for qubit states
and a much deeper ``right'' well (Fig.\ \ref{f0}), so that a qubit in the
upper state $|1\rangle$ switches by tunneling to the right-hand well with
probability close to one, whereas a qubit in the lower state $|0\rangle$
remains intact.
 Since the qubit coupling is fixed in the simple design realized
experimentally so far, the measurement is complicated by a crosstalk.
\cite{mcd05} The mechanism of the crosstalk is the following. Suppose that
one of the qubits is measured in state $|1\rangle$, which means tunneling to
the right well. The tunneling will then be followed by the dissipative
evolution (damped oscillations), eventually reaching the ground state of the
right well. These oscillations will obviously perturb the state of the
second, capacitively coupled qubit, especially because in the experiment the
oscillation frequencies in both wells are somewhat close to each other, and
both qubits are practically identical. Therefore, if the measurement of the
second qubit is made after the dissipative evolution of the first qubit, the
measurement result is likely to be wrong (it will most often be $|1\rangle$
because of the second qubit excitation \cite{mcd05}). To avoid this problem,
the measurement of both qubits should be done almost simultaneously,
\cite{mcd05} within the timescale shorter than development of the crosstalk
mechanism. However, this is not a complete solution of the problem because
the excitation of the second qubit due to crosstalk may be sufficient to
switch the qubit from the left well to the right well, even if this
excitation happens a little after the second qubit measurement, that will
also lead to the measurement error.
 We emphasize that this crosstalk mechanism is due to a fixed coupling
 between the qubits.
 It seems possible to realize schemes with adjustable coupling
in future, which will eliminate much of the measurement crosstalk discussed
here.
 Still, it is of interest to analyze the fixed-coupling case, since
that is the simplest scheme and the only one realized experimentally so far.

The present paper is devoted to the theoretical study of the measurement
crosstalk between two capacitively coupled flux-biased phase qubits. In
Sec.\ \ref{II} we study some basic properties of phase qubits. In
particular, the simplified cubic potential is discussed and the Hamiltonian
for two capacitively coupled qubits is derived.
 The measurement crosstalk is studied in Secs. \ref{III}-\ref{V}.
For definiteness, we assume that the first qubit is switched (state
$|1\rangle$), whereas the second qubit is initially in the state $|0\rangle$.
 The dynamics of the first qubit, which after the switching performs
damped oscillations in the deep well, is analyzed classically in Sec.\
\ref{III}.
 Such an approximation drastically simplifies the problem and, at
the same time, is quite accurate, since for the experimental parameters
\cite{mcd05} used here the first qubit is highly excited after measurement,
with typical quantum number over $10^2$. The second qubit in this paper is
treated both classically and quantum-mechanically.
 The classical treatment (Sec.\ \ref{IV}) involves two approaches:
the harmonic-oscillator model, \cite{mcd05} which allows for an
analytical treatment, and a numerical solution for the exact
Hamiltonian.
 In the quantum approach (Sec.\ \ref{V}), an efficient method of
numerical solution of the Schr\"odinger equation, based on using a subset of
the eigenstates of the unperturbed Hamiltonian, is developed.
 The results of the quantum approach are somewhat similar to those
of the classical treatment.
 Our quantum treatment, in contrast to the classical one, does not
allow for dissipation in the second qubit, but an insight into possible
effects of dissipation can be obtained by the comparison of the two
approaches. The conclusions following from the present work are summarized in
Sec. \ref{VI}.

\section{Flux-biased phase qubits}
\label{II}

 Before the discussion of the basic equations for coupled qubits in
Sec. \ref{IIC}, we review the properties of one qubit in Secs. \ref{IIA} and
\ref{IIB}.

\subsection{Qubit potential}
\label{IIA}

A flux-biased phase qubit schematic \cite{sim04} coincides with that of the
basic rf SQUID \cite{lik86} (Fig.\ \ref{f0}). Neglecting dissipation, it can
be described \cite{lik86} as a fictitious mechanical system with the
Hamiltonian
 \be
H=\frac{p^2}{2m}+U(\delta),
 \e{2.42}
where $\delta$ is the Josephson-junction phase difference, $p=m \dot{\delta}$
is the corresponding momentum, $m=(\Phi_0/2\pi )^2C$ is the effective mass
determined by the capacitance $C$, $\Phi_0=h/(2e)$ is the flux quantum, $e$
is the electron charge, and $U(\delta)$ is the potential energy (shown
schematically in Fig.\ \ref{f0})
 \be
U(\delta)=E_{J}\left[\frac{(\delta-\phi)^2}
{2\lambda}-\cos\delta\right].
 \e{2.43}
Here $E_{J}=\Phi_0 I_{0}/2\pi$ is the Josephson energy, $\lambda=2\pi
I_{0}L/\Phi_0$ is the dimensionless inductance, $\phi=2\pi \Phi/\Phi_0$ is
the dimensionless external magnetic flux, $I_{0}$ is the critical current,
and $L$ is the inductance.

In this and the next subsections we review the basic properties \cite{lik86}
(see also Ref.\ \onlinecite{joh05}) of the potential energy (\ref{2.43}).
Since the potential is invariant with respect to simultaneous change of
$\delta$ and $\phi$ by $2\pi n$ where $n$ is integer, we limit ourselves by
the range $0\le\phi\le2\pi$ for the external flux. Notice that the potential
is symmetric for $\phi =n\pi$ ($\Phi =n\Phi_0 /2$).

\begin{figure}[tb]
\includegraphics[width=7cm]{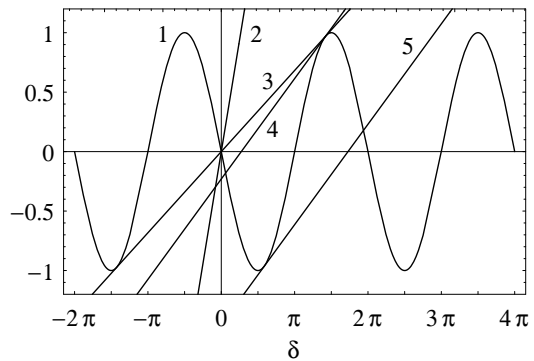}
\caption{The graphical solution of Eq.\ (\protect\ref{2.22}); see text.}
 \label{f1'}\end{figure}

The maxima and minima of the potential (\ref{2.43}) satisfy the equation
 \be
 (\delta-\phi)/\lambda=-\sin\delta ,
 \e{2.22}
which can be solved graphically.
 Figure \ref{f1'} shows the r.h.s.\ of Eq.\ (\ref{2.22}) (curve 1) and the
l.h.s.\ (straight lines 2-5) for several values of the parameters. This
equation can have only one solution when the straight-line slope is greater
than 1 (as for line 2 in Fig.\ \ref{f1'}); therefore in the case $\lambda<1$
the potential (\ref{2.43}) has only one well for any external flux $\phi$.

For $\lambda>1$ the potential may have more than one well. There will be at
most two wells if the slope of the l.h.s.\ of Eq.\ (\ref{2.22}) is greater
than the slope $1/\lambda_1$ of the line 3 in Fig.\ \ref{f1'}, which is
tangent to $-\sin\delta$ at two points $\pm\delta'$ (the line passes through
the origin because of the symmetry). This condition yields the equation
$\tan\delta'=\delta'$ with the least positive root $\delta_1'=4.493$, which
corresponds to $\lambda_1=-1/\cos\delta_1'=4.603$. Thus, for $\lambda$ in the
interval \cite{lik86}
 \be
1<\lambda<4.603
 \e{2.28}
the potential has one or two wells, depending on $\phi$.
 In particular, for the experimental parameters of Ref.\ \onlinecite{mcd05}
 used in this paper [see Eq.\ (\ref{2.16})
below] one obtains $\lambda=3.72$ (as for lines 4 and 5 in Fig.\
\ref{f1'}), which satisfies condition (\ref{2.28}).
 Similarly, one can show that the potential will have $n$ or $n+1$
wells (depending on $\phi$) if
 \be
\lambda_{n-1}<\lambda<\lambda_n\ \ (n\ge1),
 \e{2.40}
where $\lambda_0=1,\ \lambda_n=1/|\cos\delta_n'|$, and $\delta_n'$ is the
$n$th (in the increasing order) positive root of the equation
$\tan\delta'=\delta'$.
 In particular, $\lambda_2=7.790,\ \lambda_3=10.95$,
$\lambda_4=14.10$, and $\lambda_n\approx(n+1/2)\pi$ for $n\gg1$.

    The condition for a two-well potential in the case
(\ref{2.28}) can be found by considering the transition between the
one-well and two-well cases, which is illustrated by lines 4 and 5
in Fig.\ \ref{f1'}. These lines are tangent to $-\sin\delta$ at the
points $\delta_c$ (critical fluxes) which correspond to the
inflection points of the potential (\ref{2.43}) (when a well
disappears, the corresponding maximum and minimum of the potential
merge, so that both the first and second derivatives are zero at
this point). Solving the equation for the inflection points
 \be
\cos\delta_c=-1/\lambda
 \e{2.35}
(which does not depend on the external flux), we get two solutions in the
interval $(0,2\pi)$:
 \be
\delta_c=\pi/2+\arcsin(1/\lambda),\ \ \delta_c'=2\pi- \delta_c.
 \e{2.31}
Inserting these results into Eq.\ (\ref{2.22}), we finally obtain
the condition for a two-well potential:
 \begin{eqnarray}
\phi_c'<\phi<\phi_c,\label{2.29} \\
\phi_c=\pi/2+\sqrt{\lambda^2-1}+\arcsin(1/\lambda ), \,\,
\phi_c'=2\pi-\phi_c
 \label{2.32}
 \end{eqnarray}
(it is easy to show that $\phi_c'<\phi_c$ for $\lambda>1$).

For definiteness we consider the case $\pi<\phi<\phi_c$, in which the right
well is deeper than the left well.
 The depth $\Delta U_{l}$ of the left well
(i.e., the difference between the potential maximum and minimum) can be
characterized by the crude estimate of the number of discrete levels in the
well
 \be
 N_l=\frac{\Delta U_{l}}{\hbar\omega_{l}} \, ,
 \e{2.33}
where $\omega_{l}$ is the ``plasma'' frequency (classical oscillation
frequency near the well bottom) for the left well (similarly, $\omega_r$
denotes the plasma frequency for the right well):
 \be
\omega_{l,r}
 =\sqrt{E_J(1/\lambda +\cos\delta_{l,r})/m} \, ;
 \e{2.34}
here $\delta_{l,r}$ corresponds to the well minimum and obeys Eq.\
(\ref{2.22}). Notice that $N_{l,r}$ is not necessarily integer and there is
no simple relation between $N_{l,r}$ and exact number of discrete levels in
the well because of significant anharmonicity of the potential.

\begin{figure}[tb]
\includegraphics[width=7cm]{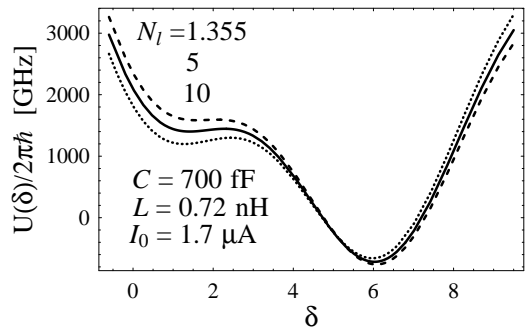}
\caption{The qubit potential $U(\delta)$ [Eq.\ (\ref{2.43})] for $N_l=10$
(dotted line), $N_l=5$ (solid line) and $N_l=1.355$ (dashed line).}
 \label{f2}\end{figure}

For numerical calculations presented in this paper we will use the following
values of the parameters from the experiment of Ref.\ \onlinecite{mcd05}:
 \be
C=700\ \mbox{fF},\ L=0.72\mbox{ nH},\ I_0=1.7\mu\mbox{A}.
 \e{2.16}
 Figure \ref{f2} shows the qubit potential $U(\delta )$ for $N_l=10$
(corresponding to $\phi=4.842$), $N_l=5$ ($\phi =5.089$), and $N_l=1.355$
($\phi =5.308$); the last value corresponds to the bias during the
measurement pulse (see below).
 The qubit levels
$|0\rangle$ and $|1\rangle$ are, respectively, the ground and the first
excited levels in the left well.

\subsection{Cubic potential}
\label{IIB}

When one of the wells is very shallow, it can be approximated by a cubic
potential. Assuming $0<\phi_c-\phi\ll\phi_c$ (shallow left well) we can
approximate $\cos \delta$ in the vicinity of $\delta_c$ as
 \bea
&& \cos\delta=\cos\delta_c\cos y-\sin\delta_c\sin
y\nonumber\\
 &&\approx-\lambda^{-1}(1-y^2/2)-\sqrt{1-\lambda^{-2}}\,(y-y^3/6),
 \ea{2.36}
where $y=\delta-\delta_c$. Then the potential (\ref{2.43}) can be
approximated by the cubic polynomial
$U_c(\delta)=\sqrt{1-\lambda^{-2}}\,E_J(\epsilon y/2-y^3/6)$, where a
constant is neglected and
 \be
\epsilon=2(\phi_c-\phi)/\sqrt{\lambda^2 -1}.
 \e{2.38}
The minimum and maximum of this potential are at
$y_l=-\sqrt{\epsilon}$ and $y_{\rm max}=\sqrt{\epsilon}$,
respectively, i.e., at
 \be
\delta_l=\delta_c-\sqrt{\epsilon}, \ \ \delta_{\rm
max}=\delta_c+\sqrt{\epsilon}.
 \e{2.44}
Shifting the axis as
$x=y+\sqrt{\epsilon}=\delta-\delta_c+\sqrt{\epsilon}$ and again
neglecting a constant, the potential can be rewritten as
 \be
U_c(\delta)=\sqrt{1-\lambda^{-2}}\,E_J(\sqrt{\epsilon}x^2/2-x^3/6).
 \e{2.39}
In this approximation \cite{lik86} the left well parameters are
 \bea
&&\omega_l=\epsilon^{1/4}(1-\lambda^{-2})^{1/4}\sqrt{E_J/m},
 \nonumber \\
&& \Delta U_l=\frac{2}{3}\, \sqrt{1-\lambda^{-2}}\,E_J\, \epsilon^{3/2},
\nonumber\\
&&N_l=\frac{2}{3\hbar}\,\epsilon^{5/4} (1-\lambda^{-2})^{1/4}\sqrt{mE_J}.
 \ea{2.41}
The validity condition for the cubic-potential description
(\ref{2.39}) is $\epsilon \ll 1$, which is well satisfied for the
left well with qubit parameters considered in the present paper.

\subsection{Two capacitively coupled qubits}
\label{IIC}

\begin{figure}[tb]
\includegraphics[width=7cm]{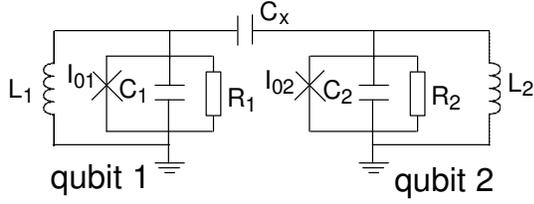}
\caption{The circuit schematic for two capacitively coupled flux-biased
phase qubits.}
 \label{f1}\end{figure}

Let us consider two capacitively coupled flux-biased phase qubits
\cite{mcd05} (Fig.\ \ref{f1}).
 The current balance for this circuit yields the equations
 \bea
 \ddot{\delta}_i+\frac{\dot{\delta}_i}{C_i'R_i}
 +\frac{2\pi I_{0i}}{\Phi_0 C_i'}\sin\delta_i+
 \frac{\delta_i-\phi_{i}}{C_i'L_i}
 =\frac{C_x}{C_i'}\ddot{\delta}_j,
 \ea{2.1}
 where qubits are numbered by $i,j=1,2\ (i\ne j)$; $C_i'=C_i+C_x$; the
 quantities
 $\delta_i,\ I_{0i},\ C_i$, and $L_i$ are respectively the
 Josephson-junction phase difference, critical current, capacitance,
 and inductance for the $i$th qubit, $C_x$ is the coupling
capacitance, and $\phi_{i}=2\pi \Phi_{i}/\Phi_0$ is the dimensionless
external magnetic flux. Dissipation in Josephson junctions is described here
using the resistively shunted junction (RSJ) model, \cite{lik86} by
introducing resistances $R_1$ and $R_2$ into the circuit (Fig.\ \ref{f1}).
For each junction the dissipation can be characterized by the energy
relaxation time
 \be
T_{1}=R_1 C_1, \,\,\, T_1'=R_2 C_2.
 \e{2.13}

In the absence of dissipation ($R_i=\infty$) Eq.\ (\ref{2.1}) can be
written in the form of Lagrange's equations, \cite{lan76}
 \be
 \frac{d}{dt}\frac{\partial L}{\partial\dot{\delta}_i}-
\frac{\partial L}{\partial\delta_i}=0,
 \e{2.11}
with the Lagrangian
 \be
 L=K-U_1(\delta_1)-U_2(\delta_2).
 \e{2.2}
 Here the $i$th qubit potential energy is [cf. Eq.\ (\ref{2.43})]
 \be
U_i(\delta_i)=E_{Ji}\left[\frac{(\delta_i-\phi_{i})^2}
{2\lambda_i}-\cos\delta_i\right],
 \e{2.3}
 where $E_{Ji}=\Phi_0 I_{0i}/2\pi$, $\lambda_i=2\pi I_{0i}L_i/\Phi_0$,
and the kinetic energy $K$ is
 \be
 K=\frac{m_1'\dot{\delta}_1^2}{2}+\frac{m_2'\dot{\delta}_2^2}{2}-
 m_x\dot{\delta}_1\dot{\delta}_2,
 \e{2.4}
 where $m_i'=(\Phi_0/2\pi)^2C_i'$ and $m_x=(\Phi_0/2\pi )^2C_x$ are the normalized
 capacitances.
 Thus, the problem of two coupled qubits is equivalent to the motion of
 a fictitious particle in the two-dimensional space $(\delta_1,\delta_2)$.
 From Eqs.\ (\ref{2.2}) and (\ref{2.4}) one can obtain the generalized
 momenta $p_i=\partial L/\partial\dot{\delta}_i$ in the
 form
  \be
 p_i=m_i'\dot{\delta}_i-m_x\dot{\delta}_j.
 \e{2.5}
It is easy to see that $p_i$ is the total (node) charge on the $i$th junction
and the adjacent coupling capacitor multiplied by $\hbar/2e$.
\cite{joh03,lik86}
 Dissipation can be accounted for by the addition \cite{lan76} of
 the friction force $F_i$ into r.h.s.\ of Eq.\ (\ref{2.11}).
 This yields Eq.\ (\ref{2.1}) if
 \be
F_i=-\frac{\Phi_0^2}{4\pi^2 R_i}\dot{\delta}_i.
 \e{2.14}

 Equations (\ref{2.5}) can be inverted, yielding
  \be
\dot{\delta}_i=p_i/m_i''+p_j/m_x' ,
 \e{2.6}
where $m_i''=m_i+(m_j^{-1}+m_x^{-1})^{-1}$ and
 $m_x'=m_1+m_2+m_1m_2/m_x$.
Inserting Eq.\ (\ref{2.6}) into (\ref{2.4}), one obtains the
Hamiltonian
 \cite{lan76} $H=K+U_1+U_2$ in the form \cite{bla03-n}
 \be
 H=\frac{p_1^2}{2m_1''}+\frac{p_2^2}{2m_2''}+\frac{p_1p_2}{m_x'}
+U_1(\delta_1)+U_2(\delta_2).
 \e{2.7}
Notice that the Hamiltonian (\ref{2.7}) can also be derived in a
direct way (without Lagrangian language) using the fact \cite{lik86}
that the node charge (muliplied by $\hbar/2e$) $p_i$ is the
conjugated variable to the phase $\delta_i$ and expressing the
combined electrostatic energy of capacitors $C_1$, $C_2$ and $C_x$
in the form of three first terms of Eq.\ (\ref{2.7}).

Instead of Lagrange's equations (\ref{2.11}), one can use Hamilton's
equations \cite{lan76} with respect to $\delta_i$ and $p_i$, viz.,
Eq.\ (\ref{2.6}) and $\dot{p}_i=-\partial H/\partial\delta_i$ or, in
view of (\ref{2.7}),
 \be
\dot{p}_i=-\frac{\partial U_i}{\partial\delta_i}.
 \e{2.12}
This equation can be extended to take dissipation into account by
adding the friction force (\ref{2.14}):
 \be
\dot{p}_i=-\frac{\partial U_i}{\partial\delta_i}-
 \frac{\Phi_0^2}{4\pi^2 R_i}\dot{\delta}_i.
 \e{2.15}

In this paper we mainly consider a system of two identical qubits, then the
subscript $i$ can be dropped in the parameter notation:
 \be
C_i=C,\ L_i=L,\ I_{0i}=I_0,\ R_i=R
 \e{2.8}
(the external flux $\phi_i$, which is the control parameter in experiments,
is assumed to be generally different for the two qubits).
 Then Eqs.\ (\ref{2.6}) and (\ref{2.7}) become
 \bea
&& H=\frac{p_1^2+p_2^2+2\zeta p_1p_2}{2(1+\zeta)m}+U_1(\delta_1)
+U_2(\delta_2),
 \label{2.9} \\
&& \dot{\delta}_i=\frac{p_i+\zeta p_j}{(1+\zeta)m},
 \ea{2.18}
where $\zeta=C_x/(C+C_x)$. Experimental parameters of Ref.\
\onlinecite{mcd05}, $C_x=6$ fF and $T_1=25$ ns, correspond to $\zeta=8.5
\times 10^{-3}$ and $R=$35.7 k$\Omega$.

The above formalism provides a basis for the analysis of the qubit-system
evolution before the measurement (the gate operation), during the measurement
pulse, and after the measurement pulse (the crosstalk).
 The latter stage is considered below.

\section{Measurement crosstalk: first qubit dynamics}
\label{III}

In the fast measurement scheme employed in Refs.\  \onlinecite{coo04} and
\onlinecite{mcd05}, a short flux pulse applied to the measured qubit
decreases the barrier between the two wells (see Fig.\ \ref{f0}), so that
the upper qubit level becomes close to the barrier top.
 In the case when level $|1\rangle$ is populated, there is a fast
population transfer (tunneling) from the left well to the right
well.
 Due to dissipation, the energy in the right well gradually
decreases, until it reaches the bottom of the right well.
 In contrast, if the qubit is in state $|0\rangle$ the
tunneling essentially does not occur.
 The qubit state in one of the two potential minima (separated by almost $\Phi_0$) is subsequently
distinguished by a nearby SQUID, which completes the measurement process.

In a system of two identical coupled qubits, crosstalk can produce
measurement error if the qubits are in different logical states. \cite{mcd05}
 For definiteness, we assume that before the measurement the qubit
system is in the state $|10\rangle$, i.e., the first qubit is in the excited
state and second qubit is in the ground state.
 Then after the measurement the first qubit performs damped
oscillations in the right well, which in the classical language \cite{lik86}
produces an oscillating (microwave) voltage $(\Phi_0/2\pi ) \,\dot\delta_1
(t)$. This voltage causes oscillating current through the coupling capacitor
$C_x$, which perturbs the second qubit. The effect is so strong that
measurement of the second qubit after the dissipative evolution of the first
qubit is practically useless: there is a little chance for the second qubit
to remain in the ground state.\cite{mcd05} The effect of crosstalk can be
significantly suppressed if the two qubits are measured practically
simultaneously (experimentally, not more than few nanoseconds apart) because
the crosstalk excitation of the second qubit takes finite time.\cite{mcd05}
Nevertheless, crosstalk leads to noticeable measurement errors even in the
case of simultaneous qubit measurement. The reason is that strong excitation
of the second qubit may lead to its switching from the left to the right
well even in absence of the measurement pulse.

    This is exactly the effect which we analyze in this paper.
We assume that the first qubit is switched to the right well at
$t=0$, while the second qubit at this time is in the ground state
and no measurement pulse is applied to the second qubit (physically,
this means that the pulse is short and does not change the qubit
state).
 Our main goal will be analysis of the measurement error,
which in this case is switching of the second qubit to the right
well due to the crosstalk excitation.

A rigorous theoretical study of the measurement crosstalk should involve a
numerical solution of quantum evolution of two coupled qubits with the
account of dissipation, which would require extensive computer resources.
 In the present paper we employ several simplified approaches, which
have the advantage of being relatively fast numerically, thus facilitating a
study of the crosstalk dependence on the parameters.
 The first qubit is always treated classically, while the second qubit is
studied both classically and quantum mechanically.

We will mainly consider two experimentally relevant cases of the second
qubit biasing, characterized by the dimensionless barrier heights $N_{l2}=5$
or 10 ($\phi_2=5.09$ and 4.84, respectively). Then the plasma frequencies are
$\omega_{l2}/2\pi =8.91$ GHz and 10.2 GHz, respectively. The crosstalk
mechanism is obviously very efficient when the first qubit oscillation is in
resonance with $\omega_{l2}$.

\vspace{0.3cm}

For the first qubit we choose the biasing parameter $N_{l1}= 1.355$
($\phi_1=5.31$) which is close to the experimental value
\cite{mcd05,rounding} at which the state $|1\rangle$ efficiently tunnels
out. The corresponding WKB tunneling rate \cite{lik81} of the state
$|1\rangle$ is $3\times10^9$ s$^{-1}$, which ensures tunneling during
few-nanosecond-long measurement pulse as in experiments of Refs.\
\onlinecite{coo04} and \onlinecite{mcd05}.
 Notice that the barrier height $N_{l1}$ is smaller
than the naive estimate 1.5 for the dimensionless energy of the state
$|1\rangle$. Actually, because of significant anharmonicity, the energies of
states $|0\rangle$ and $|1\rangle$ in this case are 0.475 and 1.26 from the
well bottom in units of $\hbar\omega_{l1}$, where $\omega_{l1}/2\pi =6.87$
GHz.
 We neglect the fact that in the experiment, after the measurement
pulse, the biasing of the left qubit returns back to $N_{l1} \sim
5$, because as we checked, this does not lead to a significant
change of the evolution dynamics in the right well.

    At the initial moment $t=0$ the first qubit is assumed to be in the right
well close to the barrier top position, with the velocity $\dot{\delta}_1=0$.
However, instead of assuming its initial energy to be the same as the energy
of state $|1\rangle$, we choose a slightly lower energy, which is below the
top of the barrier by 20\% of the well depth $\Delta U_{l1}$.
 This (somewhat
arbitrary) choice prevents unphysically slow dynamics of a classical
particle in the case when it is very close to the barrier top (quantum
dynamics due to dissipation does not significantly slow down at the energy
close to the barrier top). We have also checked that the qubit dynamics is
not too sensitive to the choice of initial energy (when it is above the left
well bottom and not too close to the barrier top).

 Since the initial energy of the first qubit with respect to the
bottom of the right well is much higher than the maximal energy of the
second qubit in the left well, one can neglect the back action of the second
qubit onto the first one while the the second qubit remains in the left well.
 Then the second term in the numerator in Eq.\ (\ref{2.18}) for the first
qubit ($i=1$) can be dropped, yielding
 \be
\dot{\delta}_1=p_1/m'',
 \e{2.23}
so Eq.\ (\ref{2.15}) with $i=1$ gives the approximate equation of
motion for the first qubit:
 \be
 \ddot{\delta}_1+\frac{\dot{\delta}_1}{C''R}
+\frac{2\pi I_{0}}{\Phi_0 C''}\sin\delta_1+\frac{\delta_1-\phi_1}{C''L}=0,
 \e{2.19}
where $C''=(1+\zeta)C$. This is obviously the usual equation for an isolated
first qubit with capacitance $C$ replaced by effective capacitance $C''=C+C_x
C/(C_x+C)$ which takes into account the series connection of the coupling
capacitance and the second junction capacitance (this corresponds to the
approximation of zero charge at the second qubit).

  Note that even though the set of equations of motion (\ref{2.1})
is equivalent to Eqs.\ (\ref{2.6}) and (\ref{2.15}), the above
approximation makes them different.
 In particular, in the case of identical qubits the equation for the
first qubit obtained from Eq.\ (\ref{2.1}) by neglecting the r.h.s.,
differs from Eq.\ (\ref{2.19}) by the substitution
$C''\leftrightarrow C'$.
 However, for small dimensionless coupling $\zeta$ (which is the experimentally relevant case assumed here),
the two equations differ by very small terms on the order of $\zeta^2$.
Physically, $C'=C+C_x$ as the effective capacitance of the first qubit
corresponds to the model in which the voltage across the second junction is
neglected (in contrast to the charge in the previous model).

\begin{figure}[tb]
\includegraphics[width=8.cm]{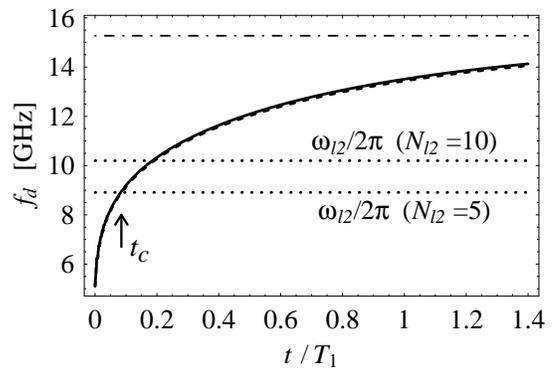}
\caption{The first-qubit oscillation frequency $f_d$ as a function of time
$t$ (normalized by the energy relaxation time $T_1$) for $C_x=0$ (solid line)
and $C_x=6$ fF (dashed line), assuming $N_{l1}=1.355$ and parameters of Eq.\
(\ref{2.16}). Dash-dotted horizontal line, $\omega_{r1}/2\pi=15.3$ GHz,
shows the long-time limit of $f_d (t)$. Two dotted horizontal lines show the
plasma frequency for the second qubit: $\omega_{l2}/2\pi =10.2$ GHz for
$N_{l2}=10$ and $\omega_{l2}/2\pi =8.91$ GHz for $N_{l2}=5$. The arrow shows
the moment $t_c$ of exact resonance in the case $N_{l2}=5$. }
 \label{f3}\end{figure}

 Equation (\ref{2.19}) shows that the first qubit performs damped
non-harmonic oscillations. Because of anharmonicity, the gradual decrease of
the qubit energy $E_1=m''\dot{\delta}_1^2/2+U_1(\delta_1)$ due to dissipation
leads to the gradual increase\cite{mcd05} of the oscillation frequency $f_d$
(driving the second qubit) which can be obtained as \cite{lan76}
 \be
f_d^{-1} (E_1)=\sqrt{2m''}\int_{a(E_1)}^{b(E_1)}
\frac{d\delta_1}{\sqrt{E_1-U_1(\delta_1)}},
 \e{2.20}
where $a$ and $b$ are the classical turning points.
 The time dependence $f_d(t)$ of the oscillation frequency
is shown in Fig.\ \ref{f3} for $C_x=0$ (solid line) and 6 fF (dashed line)
assuming $T_1=25$ ns. The curves are very close to each other showing very
small effect of the capacitance renormalization (so that even smaller
second-order effect of the difference between $C'$ and $C''$ is really
negligible). We have also checked numerically that variation of $T_1$ from
25 ns to 500 ns does not change noticeably the dependence $f_d(t)$ if the
time is normalized by $T_1$, which is rather obvious since $f_d \gg
T_1^{-1}$.

Figure \ref{f3} shows that the oscillation frequency sharply increases
initially and then slowly tends to the right-well plasma frequency
$\omega_{r1}/2\pi=15.3$ GHz (the dash-dotted horizontal line in Fig.\
\ref{f3}).
 This is explained by the fact that the initial system
energy is close to the barrier top, where the oscillation frequency is
significantly lower (it tends to zero when the energy approaches the barrier
top), while anharmonicity becomes relatively weak after the energy is no
longer close to the barrier top.

Notice that it takes a finite time $t_c$ for the first qubit dynamics to get
into resonance with the second qubit ($\omega_{l2}/2\pi$ is around 9-10 GHz,
as mentioned above); we find from Fig.\ \ref{f3} that $t_c=0.085\,T_1$ for
$N_{l2}=5$ and $t_c=0.192\,T_1$ for $N_{l2}=10$.
 As a simple estimate, this is the time after which the second qubit
becomes significantly excited.\cite{mcd05}

\section{Second qubit: Classical approach}
\label{IV}

In the approximation (\ref{2.23}), the second-qubit equation of
motion follows from Eqs.\ (\ref{2.18}) and (\ref{2.15}) with $i=2$:
 \be
\ddot{\delta}_2+\frac{\dot{\delta}_2}{(1+\zeta)T_1'} +\frac{2\pi
I_{0}}{\Phi_0 C''}\sin\delta_2
+\frac{\delta_2-\phi_2}{C''L}=\zeta\ddot{\delta}_1(t).
 \e{2.24}
Here we assume that the second-qubit relaxation time $T_1'$ may be different
from the first-qubit value $T_1$. For simplicity we will mostly neglect the
energy relaxation for the second qubit ($T_1'=\infty$), while more
experimentally relevant case $T_1'=T_1$ will be discussed only in the
classical approach.

    Equation (\ref{2.24}) has a simple physical meaning as an evolution of the
second qubit with effective junction capacitance $C''$, externally driven by
the oscillating current $(\Phi_0/2\pi )\zeta C'' \ddot{\delta}_1$. However,
considering oscillating voltage $(\Phi_0/2\pi) \dot{\delta}_1$ across the
first junction coupled to the second qubit via capacitance $C_x$, one would
expect the driving current to be $(\Phi_0/2\pi )C_x \ddot{\delta}_1$ [as in
Eq.\ (\ref{2.1})]. The relative difference between $\zeta C''$ and $C_x$ is
on the order of $\zeta^2\ll 1$, which is the accuracy of treating two qubits
separately.

    In this section we consider classical dynamics of the second qubit due to
the oscillating driving force $\zeta\ddot{\delta}_1(t)$ in Eq.\ (\ref{2.24})
with a slowly varying period and amplitude.

\subsection{Harmonic-oscillator model}

Though Eq.\ (\ref{2.24}) is simpler than the exact equations
(\ref{2.18}) and (\ref{2.15}), the behavior described by Eq.\
(\ref{2.24}) is still complicated and generally chaotic.
\cite{gut90}
 To get an insight into the behavior of the second qubit, we first
model it by a harmonic oscillator, \cite{mcd05} i.e., substitute
Eq.\ (\ref{2.24}) by
 \be
\ddot{x}+\omega_{l2}^2x=\zeta\ddot{\delta}_1(t),
 \e{3.1}
where $x=\delta_2-\delta_{l2}$, $\delta_{l2}$ is the left-well minimum
position, and $\omega_{l2}$ is the unperturbed plasma frequency. Actually,
the small-vibration frequency in Eq.\ (\ref{2.24}) should be different from
$\omega_{l2}$ due to the mass (capacitance) renormalization $m\rightarrow
m''$, so that $\omega_{l2} \rightarrow \omega_{l2}/\sqrt{1+\zeta}$; however,
for small coupling considered here ($\zeta\alt1$\%) we neglect the
difference.
 Correspondingly, the oscillator energy is
$E_2=m(\dot{x}^2+\omega_{l2}^2x^2)/2$ (here we also neglect the difference
between $m$ and $m''$).

\begin{figure}[tb]
\includegraphics[width=9cm]{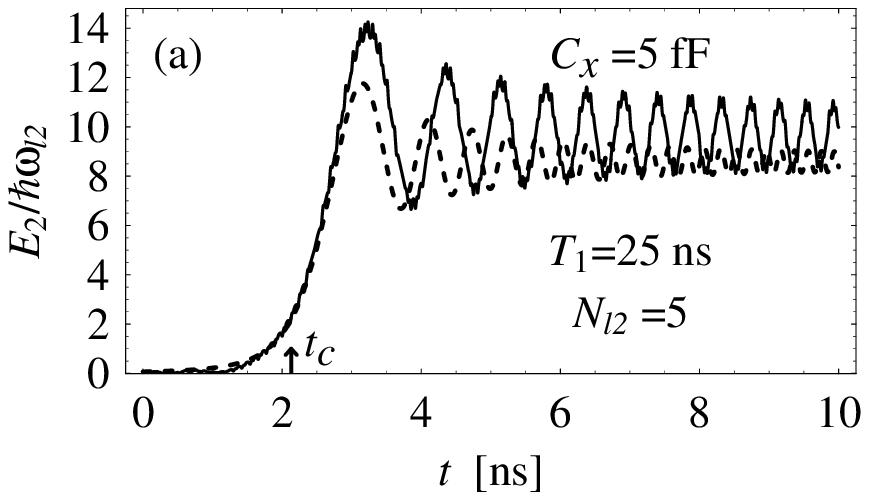}\\
\vspace{.2cm}
\includegraphics[width=9cm]{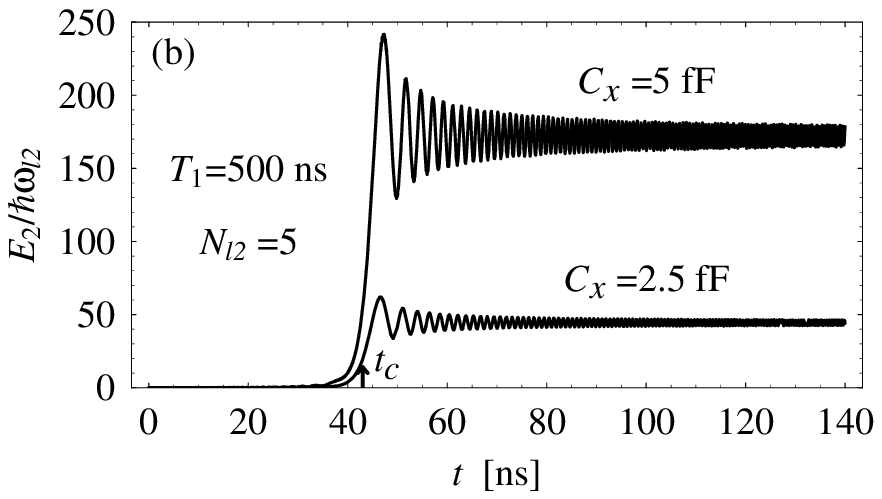}
\caption{The second qubit energy $E_2$ (in units of
$\hbar\omega_{l2}$) in the oscillator model as a function of time
$t$ (in ns) for (a) $C_x=5$ fF and $T_1=25$ ns and (b) $C_x=2.5$ fF
and 5 fF and $T_1=500$ ns, while $N_{l2}=5$.
 Dashed line in (a) shows approximation using Eq.\ (\ref{3.10}).
 The arrows show the moment $t_c$ when the driving frequency $f_d$ (see Fig.\
\protect\ref{f3}) is in resonance with $\omega_{l2}/2\pi = 8.91$ GHz. }
 \label{f4}\end{figure}

We have simulated Eq.\ (\ref{3.1}) numerically, assuming the system
initially at rest at the potential minimum, $x(0)=\dot{x}(0)=0$, and
checked that the oscillator energy coincides with the analytical
solution: \cite{lan76}
 \be
E_2(t)=\frac{\zeta^2m}{2}\left|\int_0^tdt'
e^{-i\omega_{l2}t'}\ddot{\delta}_1(t')\right|^2.
 \e{3.3}
Figure\ \ref{f4} shows the time dependence of the energy $E_2(t)$ in units
of $\hbar\omega_{l2}$ for $\omega_{l2}/2\pi =8.91$ GHz, which corresponds to
$N_{l2}=5$ (parameters of the first qubit evolution have been discussed in
Sec.\ III and correspond to Fig.\  \ref{f3}). One can see that the energy
$E_2$ remains very low until a sharp increase followed by gradually
decreasing oscillations. This behavior can be easily explained by changing
in time frequency $f_d$ of the driving force (Fig.\ \ref{f3}) which passes
through the resonance with the second qubit. \cite{mcd05}

   For an analytical analysis, let us consider the vicinity of the moment
$t_c$ of exact resonance, $\omega_d(t_c)=\omega_{l2}$ (here $\omega_d=2\pi
f_d$), and approximate $\ddot{\delta}_1(t)$ as a harmonic signal
$\ddot\delta_1(t)=A(t) \exp(i\int^t\omega_d\, dt)$ with constant amplitude
$A=A(t_c)$ and linearly varying frequency
 \be
\omega_d(t)=\omega_{l2}+\alpha(t-t_c),
 \e{3.5}
with $\alpha=\dot{\omega}_d(t_c)>0$ (we have neglected the
complex-conjugated term and higher-order harmonics as being out of
resonance).
 Thus the problem is reduced to passage
of a harmonic oscillator through resonance with a constant
rate.\cite{lew32,abl73} Also assuming slow crossing, $\alpha \ll
\omega_{l2}^2$, and shifting the lower endpoint of integration in
Eq.\ (\ref{3.3}) to $-\infty$ (which is a good approximation for
$t_c\gg\sqrt{2/\alpha}$), we obtain
   \begin{eqnarray}
E_2(t)= E_0 F(\tilde{t}), \,\,\, E_0=\frac{\pi\zeta^2m A^2}{\alpha} , \,\,\,
\tilde{t}=\frac{t-t_c}{\sqrt{2/\alpha}} ,
\nonumber\\
F(\tilde{t})=\frac{1}{\pi}\left|\int_{-\infty}^{\tilde{t}}
e^{i\eta^2} d\eta\right|^2
 =\frac{1}{4}\left|1+{\rm erf}\left(\frac{\tilde{t}}{\sqrt{i}}\right)
 \right|^2.
 \label{3.10}
 \end{eqnarray}
Notice that the function $F(\tilde{t})$ with $\tilde{t}$ proportional to a
spatial coordinate describes the Fresnel diffraction \cite{lan75} and has
the following asymptotic dependence:
 \bes{3.11}
 \bea
&&F(\tilde{t})\approx1+\frac{\sin(\tilde{t}^2-\pi/4)}{\sqrt{\pi}\,\tilde{t}}\
\ \ \ \mbox{for} \ \tilde{t}\gg 1,
\label{3.11a}\\
&&F(\tilde{t})\approx (4\pi \tilde{t}^2)^{-1} \ \ \ \ \mbox{for} \
-\tilde{t}\gg 1.
 \ea{3.11b}
 \ese
 The oscillating term in Eq.\ (\ref{3.11a}) describes the beating
between the oscillator and driving force frequencies, with the difference
frequency increasing in time, $d(\tilde{t}^2)/dt=\alpha(t-t_c)$, and
amplitude of beating decreasing as $1/\tilde{t}$ (see dashed line in Fig.\
\ref{f4}a). Notice that $F(0)=1/4$, $F(\infty )=1$, and the maximum value is
$F(1.53)=1.370$, so that $E_0$ is the long-time limit of the oscillator
energy $E_2$, while the maximum energy is 1.37 times larger:
 \be
E_{\rm 2,max}=\frac{1.37\pi\zeta^2m A^2}{\alpha}.
 \e{3.12}

     This result for the energy can be understood in the following way.
In case of exact resonance, the oscillation amplitude of $x(t)$ in Eq.\
(\ref{3.1}) increases linearly in time with the rate \cite{lan76} $\zeta
A/\omega_{l2}$. The effective time of resonance $\Delta t$ corresponds to a
significant phase shift due to beating: $\alpha (\Delta t)^2\sim 1$.
Therefore the resulting amplitude after the resonance crossing is $\sim
\zeta A/\omega_{l2}\sqrt{\alpha}$ and the corresponding energy is $E_{2}
\sim \zeta^2m A^2/\alpha$, which coincides with Eq.\ (\ref{3.12}) up to a
numerical factor.

For the parameters of Fig.\ \ref{f4} ($N_{l2}=5$) we find from Fig.\
\ref{f3} that exact resonance between $f_d(t)$ and $\omega_{l2}/2\pi =8.91$
GHz occurs at $t_c=0.085\,T_1$ (in particular, $t_c=2.13$ ns for $T_1=25$ ns
and $t_c=43$ ns for $T_1=500$ ns) and $\alpha=110\, \mbox{ns}^{-1}/T_1$.
 To compare Eq.\
(\ref{3.10}) with the numerical results, we also need the value of $A$. It
can be estimated as $A=\omega_{l2}^2 \tilde{A}/2$, where $\tilde{A}$ is the
amplitude of $\delta_1(t)$ oscillations at $t=t_c$ (the factor 1/2 comes
from our definition of $A$ as half of the amplitude of $\ddot \delta_1$
oscillations). Using the numerical result $\tilde{A}\simeq 2.7$
corresponding to the resonance with the second qubit at $N_{l2}=5$, we find
$A=4.3\times10^3$ ns$^{-2}$ (this resonance happens at 121 GHz below the
barrier top; $\tilde{A}$ is defined as half of the full span of $\delta_1$
oscillations -- see Fig.\ \ref{f2}).
  The dashed line in Fig.\ \ref{f4}(a) shows the corresponding
analytical result (\ref{3.12}).  One can see that the analytics fits the
oscillator energy at $t=t_c$ pretty well; however, the maximum energy
$E_{\rm 2,max}$ given by Eq.\ (\ref{3.12}) is somewhat different from the
numerical result: $E_{\rm 2,max}/\hbar\omega_{l2}=11.8$ versus 14.2
numerically. (The difference decreases with increase of $T_1$: for $C_x=5$
fF and $T_1=500$ ns the corresponding numbers are 235 and 242.)
 A noticeable discrepancy between the analytical and numerical
results in Fig.\ \ref{f4}(a)  can be attributed mainly to the fact that $A$
and $\alpha$ change with time, in contrast to the assumptions made in the
derivation of Eq.\ (\ref{3.10}). It is interesting to notice that $A(t)$
initially increases because of the frequency $f_d$ increase, while it starts
to decrease at $t>0.52 \, T_1$ (after reaching the maximum of $5.8\times10^3$
ns$^{-2}$) because of $\tilde{A}$ decrease. Actually, a good fit of the
numerical results by the values of $A$ obtained as $A=(2\pi f_d)^2
\tilde{A}/2$ is quite surprising, because oscillations $\ddot\delta_1(t)$ are
strongly non-harmonic (even having three maxima and three minima per period).
We have also calculated $A$ using some advanced analysis of $\ddot\delta_1
(t)$ and found values very close to the simple estimate. (This other method
is based on calculating Fourier transform of $\ddot\delta_1$ within a time
interval around $t_c$, cutting off the spectrum above the minimum between the
first and second Fourier peaks, calculating inverse Fourier transform, and
finding the oscillation amplitude at $t=t_c$.)

Increase of $N_{l2}$ leads to more efficient excitation of the second qubit.
For example, for $N_{l2}=10$ and other parameters as in Fig.\ \ref{f4}(a),
the numerical maximum value of $E_{\rm 2,max}/\hbar\omega_{l2}$ becomes 30.5
(more than twice larger compared to the case $N_{l2}=5$). This happens
because of the decrease of $\alpha (t)$ and increase of $A(t)$ with time, and
correspondingly with $N_{l2}$. [For $N_{l2}=10$ (so that $\omega_{l2}/2\pi
=10.2$ GHz), we find $t_c=0.192\,T_1$ (i.e., $t_c=4.8$ ns for $T_1=25$ ns
and $t_c=96$ ns for $T_1=500$ ns), $\alpha=57\,\mbox{ns}^{-1}/T_1$, and
 $A(t_c)=5.2\times 10^{3}$ ns$^{-2}$.]

    So far we have completely neglected the energy relaxation in
the second qubit. Since the effective time of resonance is $\Delta t\sim
3/\sqrt{\alpha}$ [rise time of the function $F(\tilde{t})$ from the 10\%
level to the maximum], the neglected effect should not be important (less
than $\sim$10\%) for $T_1\agt 30/\sqrt{\alpha}$. Using the estimate
$\alpha=110\, \mbox{ns}^{-1}/T_1$ (see above), we find that taking into
account the second qubit relaxation would not change significantly our
results for $E_{\rm 2,max}$ if $T_1\agt$ 10 ns, which justifies our model.

Let us discuss the dependence of the maximum energy $E_{\rm 2,max}$
of the second qubit on $C_x$ and $T_1$. Taking into account that
$\zeta\propto C_x$ (for $C_x\ll C$) and $\alpha\propto1/T_1$, we
obtain from Eq.\ (\ref{3.12}) the scaling
 \be
E_{\rm 2,max}\propto C_x^2T_1.
 \e{3.16}
As seen from Fig.\ \ref{f4}, numerical results confirm the obvious
scaling $E_{\rm 2,max}\propto C_x^2$, while the scaling $E_{\rm
2,max}\propto T_1$ is not very accurate, but is still good as a
first approximation.

In this subsection we have treated the second qubit as a harmonic
oscillator. However, to analyze the measurement error due to the crosstalk,
we have to assume switching from the left well to the right well when
$E_{\rm 2,max}> N_{l2}\hbar\omega_{l2}$ (which is surely not fully
consistent with the oscillator model). All curves in Fig.\ \ref{f4}
($N_{l2}=5$) correspond to such switching, leading to the measurement error.
 The measurement error can be improved by decreasing the coupling
capacitance $C_x$, which should be chosen to be smaller for larger
$T_1$.
 Equation (\ref{3.16}) implies that to avoid the errors due to crosstalk,
one needs to choose
    \be
C_x < C_{x,T}=B/\sqrt{T_1},
    \e{bound1}
where $C_{x,T}$ is the threshold coupling capacitance.
 From the numerical simulations (see
Fig.\ \ref{f4}) we obtain $B\simeq 15$ fF$\sqrt{\mbox{ns}}$ in the case
$N_{l2}=5$.
 For $N_{l2}=10$ we get $B\simeq 14$ fF$\sqrt{\mbox{ns}}$, yielding  a little stricter
condition than for $N_{l2}=5$.
 Notice that for experimental
parameters of Ref.\ \onlinecite{mcd05} ($C_x$=6 fF, $T_1=$25 fF) this bound
is exceeded approximately twice, which is an indication that our simple model
is not sufficiently accurate. As we will see in the next subsection, the
theoretical bound is softer (higher) when we use actual potential profile
for the second qubit instead of using the harmonic oscillator model.

\subsection{Actual qubit potential}
 \label{IVB}

\begin{figure}[tb]
\includegraphics[width=9cm]{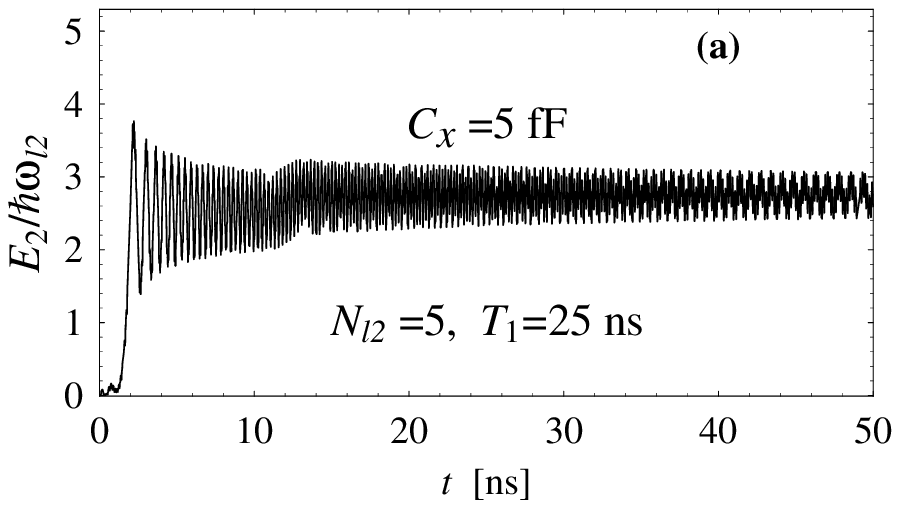}\\
\vspace{.2cm}
\includegraphics[width=9cm]{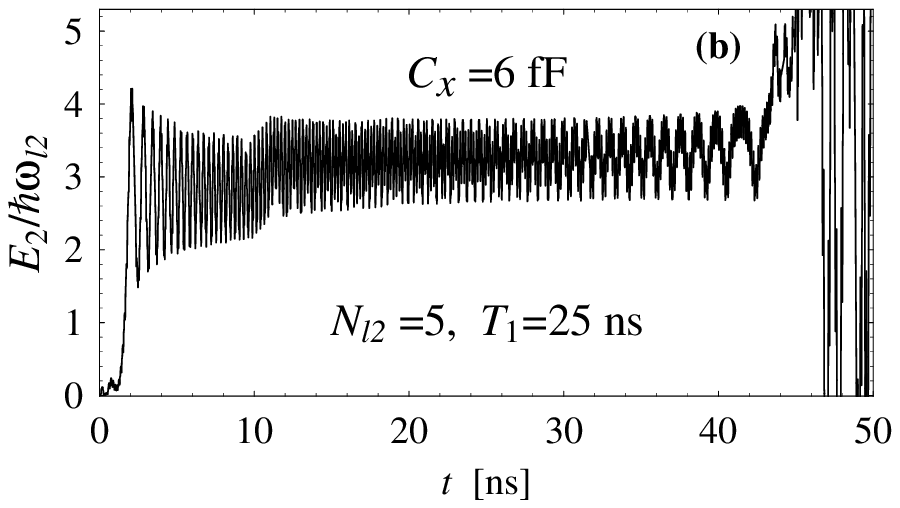}\\
\vspace{.2cm}
\includegraphics[width=9cm]{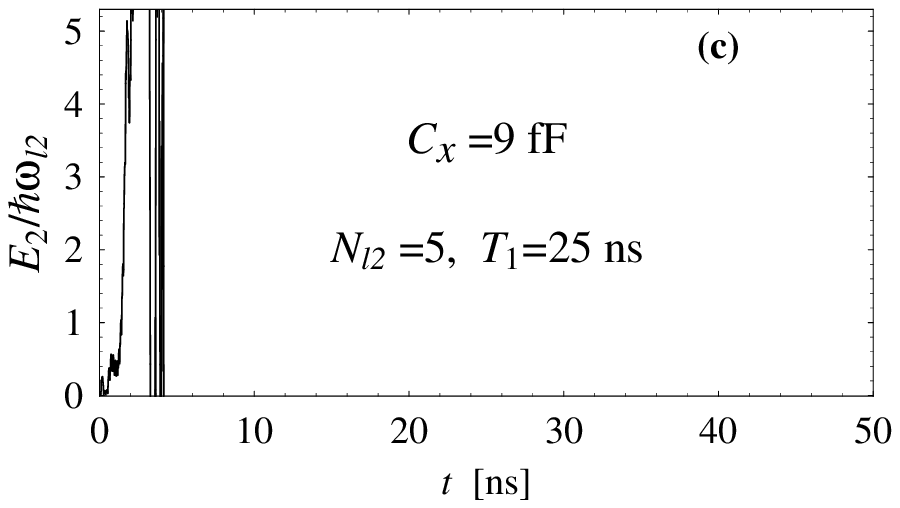}
\caption{The second-qubit energy $E_2$ (in units of $\hbar\omega_{l2}$) as a
function of time $t$ (in ns) for $N_{l2}=5$, $T_1=25$ ns, and (a) $C_x=5$
fF, (b) $C_x=6$ fF, and (c) $C_x=9$ fF. The classical model with actual
qubit potential is used; energy relaxation in the second qubit is neglected,
$T_1'=\infty$. In (b) and (c) the qubit switches (goes over the barrier) at
44 ns and 2.1 ns, respectively.}
 \label{f7}\end{figure}

\begin{figure}[tb]
\includegraphics[width=9cm]{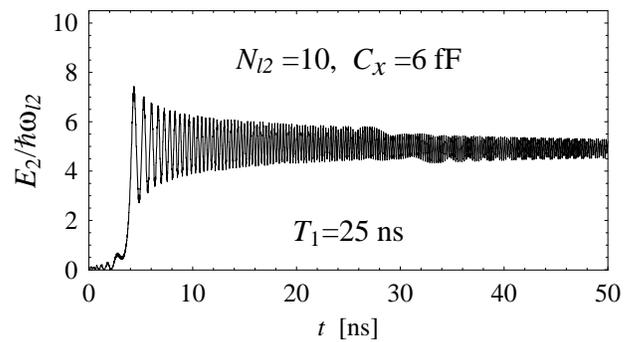}
\caption{Same as in Fig.\ \protect\ref{f7} for $N_{l2}=10$, $T_1=25$ ns, and
$C_x=6$ fF.}
 \label{f7'}\end{figure}

Let us analyze the second qubit evolution still using a classical
model, but taking into account the actual potential profile
$U_2(\delta_2)$, i.e., solving Eq.\ (\ref{2.24}) instead of the
simplified equation (\ref{3.1}).
 Figures \ref{f7} and \ref{f7'} show the time dependence of the
second-qubit energy,
 \be
 E_2=m''\dot{\delta_2}^2/2+U_2(\delta_2),
 \e{3.17}
in the absence of dissipation in the second qubit ($T_1'=\infty$) for
$N_{l2}=5$ and 10, while $T_1=25$ ns. (In this subsection  we take into
account the mass renormalization $m\rightarrow m''$ explicitly, even though
this does not lead to a noticeable change of results.)
 A comparison of Figs.\ \ref{f4}(a) and \ref{f7} shows that in both models
the qubit energy remains small before a sharp increase in energy. However,
there are significant differences due to account of anharmonicity:
 (a) The sharp energy increase occurs earlier than in the oscillator model
(the position of short-time energy maximum is shifted approximately from 3 ns
to 2 ns);
 (b) The excitation of the qubit may be to a much lower energy than for the
 oscillator;
 (c) After the sharp increase, the energy occasionally undergoes
noticeable upward (as well as downward) jumps, which may overshoot the
initial energy maximum;
 (d) The model now explicitly describes the qubit escape (switching) to the right
well [Figs.\ \ref{f7}(b) and \ref{f7}(c)]; in contrast to the oscillator
model, the escape may happen much later than initial energy increase; for
example, in Fig.\ \ref{f7}(b) the escape happens at $t\simeq 44$ ns $\gg t_c
\simeq 2.1$ ns.

    The properties (a) and (b) can be understood by taking into account the fact
that the oscillation frequency in the second qubit {\it decreases}
with the energy increase (it should become formally zero at the top
of the barrier), while the driving frequency {\it increases} with
time (Fig.\ \ref{f3}).
 Therefore, initially small out-of-resonance beatings when
$\omega_d<\omega_{l2}$ are amplified because of the positive
feedback: larger amplitude makes it closer to the resonance, which
increases the amplitude even more. This makes the non-excited state
unstable, which leads to a sharp increase of the qubit energy
earlier then the condition $\omega_d=\omega_{l2}$ is satisfied.
 The same mechanism is also responsible for lower qubit
excitation, when compared to the harmonic oscillator model:
 the resonance cannot be as efficient as in the harmonic oscillator
model since the qubit excitation quickly moves the qubit frequency out of the
resonance.
 The property (c) is related to crossing of higher-order
resonances, which occur when $\omega_d(t)$ is commensurate \cite{lan76} with
the oscillation frequency of the system, which itself depends on the energy
$E_2(t)$ and hence on the time. Similar mechanism is responsible for the
qubit switching at $t \gg t_c$; in particular, in Fig.\ \ref{f7}(b) the
switching happens when the driving frequency $f_d$ becomes approximately
twice larger than the second qubit frequency.

In contrast to the oscillator model, Eq.\ (\ref{2.24}) for the actual qubit
potential cannot be solved analytically, \cite{note2} so we rely only on the
numerical simulations. We are interested in the conditions, for which the
system remains in the left well. Generally, the qubit excitation increases
with increase of the coupling $C_x$; therefore one expects a certain
threshold value $C_{x,T}$ (depending on $T_1$ and other parameters), which
separates the switching and no-switching scenarios. However, because of the
complex dynamics of the system, the dependence on $C_x$ is non-monotonous,
so that increasing $C_x$ may sometimes change switching case into
no-switching case. In this situation, we define $C_{x,T}$ as a minimum value
at which the switching happens (even though larger $C_x$ may correspond to
no-switching). Similar to the harmonic oscillator model, we expect that
$C_{x,T}$ generally decreases with increase of $T_1$; however, because of
the complex dynamics, the dependence $C_{x,T}(T_1)$ should not necessarily
be monotonous.

\begin{figure}[tb]
\includegraphics[width=8cm]{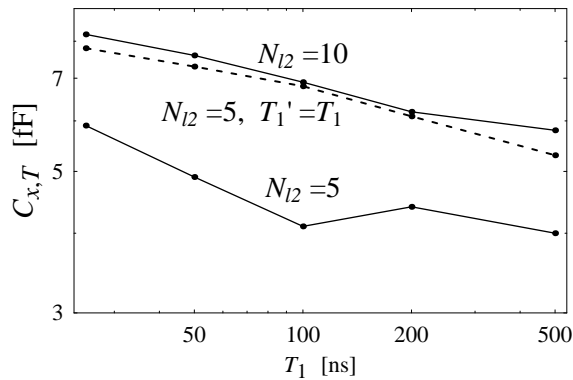}
\caption{The log-log plot of the threshold coupling capacitance $C_{x,T}$
vs.\ $T_1$ in the classical model assuming $T_1'=\infty$ (solid lines) or
$T_1'=T_1$ (dashed line) for $N_{l2}=5$ (dashed and lower solid line) and 10
(upper solid line). The crosstalk excitation does not switch the second
qubit if $C_x<C_{x,T}$.
 The numerical data are shown by the dots; the lines are just guides for the eye.}
 \label{f8}\end{figure}

The dots connected by two solid lines in Fig.\ \ref{f8} show the numerically
calculated $C_{x,T}$ for $N_{l2}=5$ and 10, for 5 values of $T_1$ ranging
from 25 ns to 500 ns (so far we still assume $T_1'=\infty$). For these
calculations we have used the increment of 0.1 fF for $C_x$ and simulated the
qubit dynamics in the time interval $0\le t\le6T_1$. Notice that the lines
are not smooth (the lower line even has a bump), which is the result of
irregular nonlinear dynamics of the system.
 Nevertheless, the numerical results confirm the generally decreasing dependence
$C_{x,T}(T_1)$, and fitting solid lines by the formula
 \be
C_{x,T}(T_1)\simeq B\,T_1^{-\beta}
 \e{3.18}
(where $C_{x,T}$ is measured in fF while $T_1$ is measured in ns) we obtain
$B\simeq 8$ and 12 for $N_{l2}=5$ and 10, respectively, while $\beta\simeq
0.12$ for the both lines.

 Since the gate speed is proportional to $C_x$, it is
advantageous to have higher $C_x$.
 The above results show that raising the barrier after the
measurement pulse to $N_{l2}=10$ would allow us to increase $C_x$ in
comparison with the case $N_{l2}=5$.
  The reason for this can be understood by comparing Figs.
\ref{f7}(b) and \ref{f7'}, which show that in the case $N_{l2}=10$  the
sharp energy increase is lower relative to the barrier top than for
$N_{l2}=5$.
 Note that the dependence (\ref{3.18}) is much weaker than the relation
$C_{x,T}\propto T_1^{-1/2}$ obtained in the oscillator model, which is
advantageous for design of qubits with weak decoherence (large $T_1$).

\begin{figure}[tb]
\includegraphics[width=9cm]{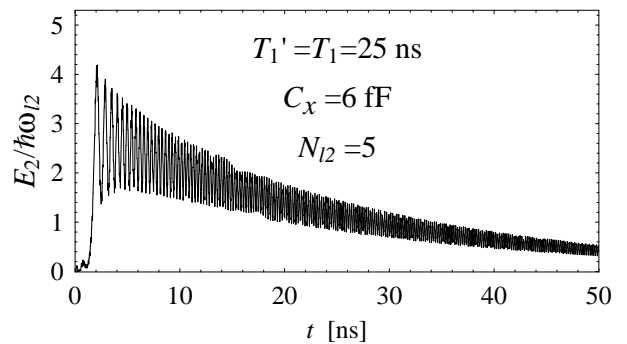}
\caption{The second-qubit energy $E_2(t)$ in the classical model taking into
account energy dissipation in the second qubit, for $N_{l2}=5$, $C_x=6$ fF,
and $T_1'=T_1=25$ ns. [Compare with Fig.\ \ref{f7}(b).] }
 \label{f9}\end{figure}

Now let us consider the effect of dissipation in the second qubit. The
dashed line in Fig.\ \ref{f8} shows $C_{x,T}(T_1)$ dependence in the
presence of dissipation ($T_1'=T_1$) for the same other parameters as for
the lower solid line (for which $T_1'=\infty$). As we see, account of
dissipation increases $C_{x,T}$ quite noticeably, which contradicts the
conclusion from the harmonic oscillator model (predicting no significant
dependence). The reason is that for $C_x$ slightly above $C_{x,T}$ the
switching in the model without dissipation usually occurs significantly
later than the initial sharp increase of the energy [see Fig.\ \ref{f7}(b)]
and is caused by ``secondary'' jumps of the energy due to strongly nonlinear
dynamics, as discussed above. Dissipation in the second qubit (Fig.\
\ref{f9}) shortens significantly the time interval during which the
switching due to secondary jumps is possible, thus increasing $C_{x,T}$. (We
would also like to mention a possibility of a system return into the left
well after the escape into the right well, which may take place with or
without dissipation.) Fitting the dashed line in Fig.\ \ref{f8} by the
power-law dependence (\ref{3.18}), we find $B \simeq 12$ and $\beta \simeq
0.13$, so that the scaling power $\beta$ is practically the same as in the
no-dissipation case, while $B$ becomes considerably larger.

\section{Second qubit: Quantum approach}
\label{V}

In the quantum approach the second qubit is described by the wave function
$\Psi(\delta,t)$ (in this section we often omit the subscript 2 to shorten
notations), which obeys the Schr\"odinger equation
 \be
i\hbar\frac{\partial\Psi}{\partial t}=H(t)\Psi.
 \e{3.19}
Here the Hamiltonian
 \be
H(t)=\frac{\hat{p}^2+2\zeta p_1(t)\hat{p}}{2(1+\zeta)m}+U(\delta),
 \e{3.20}
in which $\hat{p}=-i\hbar (\partial/\partial\delta )$, follows from Eq.\
(\ref{2.9}) by considering $p_2$ as the operator $\hat{p}$, while $p_1 (t)$
and $\delta_1(t)$ are considered as classical functions of time obtained from
Eqs.\ (\ref{2.19}) and (\ref{2.23}); the first-qubit energy in this case
does not contribute to the Hamiltonian (\ref{3.20}).

The term linear in $\hat{p}$ in Eq.\ (\ref{3.20}) has the same form as for
the interaction of a charged particle with a time-dependent electric field
described by a vector potential. \cite{lan75}
 Using the gauge transformation \cite{lan77}
 \be
\Psi(\delta,t)=\Psi'(\delta,t)e^{-i\zeta p_1(t)\delta/\hbar},
 \e{3.21}
we can replace the vector-potential by a scalar potential in the
Hamiltonian.
 Then Eq.\ (\ref{3.19}) becomes
 \be
i\hbar\frac{\partial\Psi'}{\partial t}=H'(t)\Psi',
 \e{3.22}
in which the Hamiltonian [subtracting $c$-number term
$\zeta^2p_1^2(t)/(2m'')$ and using Eq.\ (\ref{2.23})] becomes
 \bea
 &&H'(t)=H_0+V(t),\nonumber\\
&&H_0=\frac{\hat{p}^2}{2m''}+U(\delta),\ \
 V(t)=-\zeta m''\ddot{\delta}_1(t)\, \delta.
 \ea{3.23}
This Hamiltonian exactly corresponds to the classical model used in the
previous section. \cite{rem-gauge} Similarly to the classical case, the
difference between $\zeta m''$ and $m_x=(\Phi_0/2\pi )^2 C_x$ in the formula
for $V(t)$ should not be taken seriously (as being within the accuracy of
treating two qubits separately).

The partial differential equation (\ref{3.22}) can be reduced to an
infinite set of ordinary differential equations, \cite{lan77} using
the expansion of the wavefunction over the eigenfunctions
$\psi_n(\delta)$ of $H_0$,
 \be
\Psi'(\delta,t)=\sum\nolimits_n a_n(t)\, \psi_n(\delta).
 \e{3.24}
 Inserting expansion (\ref{3.24}) into Eq.\ (\ref{3.22}) yields the set of
 equations for the coefficients $a_n(t)$:
 \be
\dot{a}_n=-i(E_n/\hbar)a_n+
 \frac{i\zeta m''\ddot{\delta}_1(t)}{\hbar}
 \sum\nolimits_{n'}\delta_{nn'}a_{n'},
 \e{3.25}
where $E_n$ is an eigenvalue of $H_0$ and $\delta_{nn'}$ is the ``position''
matrix element:
 \be
\delta_{nn'}=\int_{-\infty}^\infty\psi_n^*(\delta) \, \delta \,
\psi_{n'}(\delta)\, d\delta
 \e{3.26}
(notice that we will use a different notation for the Kronecker symbol).

We have calculated the eigenstates and eigenvalues of $H_0$ numerically,
using the Fourier grid Hamiltonian method \cite{Marston} (same as periodic
pseudospectral method \cite{for96}). After obtaining eigenfunctions, we have
calculated the matrix $\delta_{nn'}$ (\ref{3.26}) and solved numerically
Eqs.\ (\ref{3.25}), restricting the space to a finite subset of the states.
 For a given $N_{l2}$, we take a reasonably small number $n_r$ of
consecutive states ($n=n_i,n_i+1,\dots,n_i+n_r$), which include all
left-well states and provide a sufficiently good approximation to the exact
solution (the choice of the subset of states is discussed below).
 The column vector $a=(a_{n_i},\dots,a_{n_i+n_r})^T$ satisfies
the equation [cf. Eq.\ (\ref{3.25})]
 \be
i\hbar\dot{a}=\tilde{H}(t)\, a ,
 \e{3.31}
in which the matrix
 \be
\tilde{H}_{nn'}(t)=E_n\delta^K_{nn'}-
 \zeta m''\ddot{\delta}_1(t)\delta_{nn'}\ \
 (n_i\le n,n'\le n_i+n_r),
 \e{3.30}
is the Hamiltonian of the system in the restricted Hilbert space spanned by
the subset of states (here $\delta^K_{nn'}$ is the Kronecker symbol).

We define the probability $P_l(t)$ to find the second qubit in the left well
as
 \be
P_l(t)=\sum\nolimits_{n_l}P_{n_l}(t),\ \
 P_n(t)=|a_n(t)|^2,
 \e{3.27}
where $P_n(t)$ is the probability of state $n$ occupation, and the summation
is only over the states localized in the left well.
 We will also denote $P_{n_l}(t)$ as $Q_k(t)$, where $k$
enumerates the states in the left well, starting from $k=0$ (the left-well
ground state).
 We define the switching probability as $P_s(t)=1-P_l(t)$.
Notice that we consider transition to delocalized states (above the barrier)
as escape from the left well (even though in this case there is a
possibility of ``repopulation'' of the left well if dissipation is taken
into account).

\begin{figure}[tb]
\includegraphics[width=9cm]{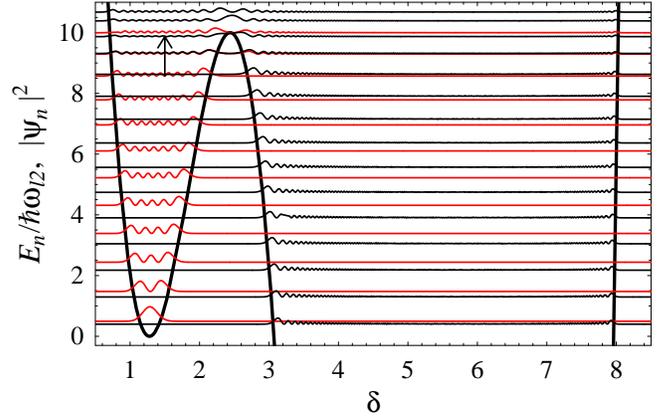}
\caption{The second qubit potential $U(\delta)$ (thick line) and
eigenfunctions $|\psi_n(\delta)|^2\ (145\le n\le171)$, shifted vertically by
the energy eigenvalues $E_n$ (in units of $\hbar\omega_{l2}$) for $N_{l2}=10$
and $C_x=6$ fF ($\omega_{l2}/2\pi=10.2$ GHz).
 The energy origin is chosen at the minimum of the left well. An arbitrary scale
for $|\psi_n|^2$ is chosen to be 1/8 of the $E_n/\hbar\omega_{l2}$ scale. The
arrow illustrates the dominating transition responsible for the escape from
the level $k=9$ at $t\simeq 16$ ns. }
 \label{f10}\end{figure}

 Figure \ref{f10} shows (for the case $N_{l2}=10$) the eigenfunctions
$|\psi_n(\delta)|^2$ and the corresponding energies $E_n$ for $145\le
n\le171$, where the level numbering starts with $n=0$ for the ground state
(in the right well).
 One can distinguish 3 types of states: (a) 12 states localized in
the left well ($n=146,148,\dots,166,169$ or, respectively, $k=0,\dots,11$),
(b) delocalized states ($n\ge170$), and (c) states localized in the right
well (the remaining states). Depending on the barrier height $N_{l2}$
(controlled by the external flux $\phi_2$),  the resonant states may also be
present: when the energies of states localized in the left and right wells
approach each other sufficiently close, the states mix and become
delocalized.
 Actually, in Fig.\ \ref{f10} the left-well states $k=10$ and 11 are
partially delocalized due to interaction (tunneling) with neigboring
right-well states. Notice that even though the left-well energies are
practically insensitive to the coupling capacitance $C_x$ ($C_x=6$ fF for
Fig.\ \ref{f10}), their relative energy shift with the right-well level comb
depends on $C_x$ significantly because the right well is very deep, $n\agt
10^2$.

\begin{figure}[tb]
\includegraphics[width=9cm]{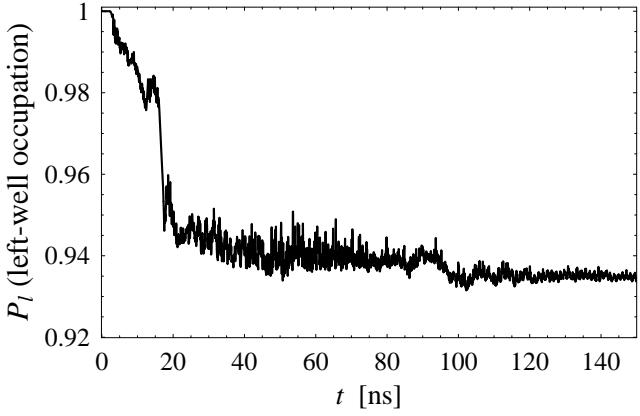}
\caption{The probability $P_l=1-P_s$ for the second qubit to remain in the
left well as a function of time $t$ for $N_{l2}=10$, $C_x=6$ fF, and
$T_1=25$ ns. }
 \label{f11}\end{figure}

\begin{figure}[tb]
\includegraphics[width=8cm]{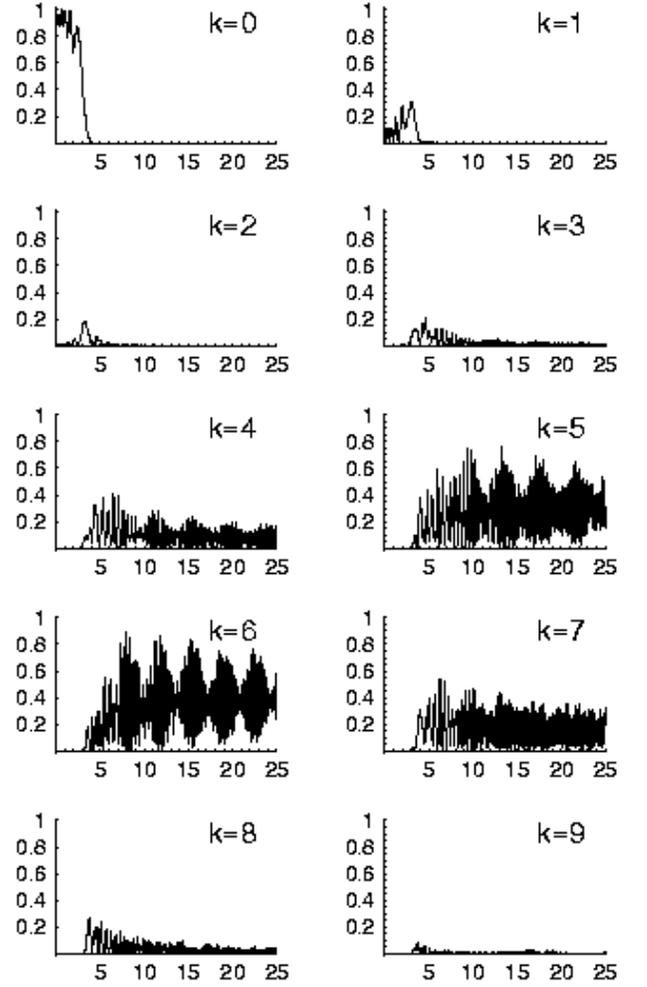}
\caption{The occupation probabilities $Q_k(t)$ for the left-well
levels $k$ as functions of time $t$ (in ns) for $N_{l}=10$, $C_x=6$
fF, and $T_1=25$ ns. }
 \label{f12}\end{figure}

Figure \ref{f11} shows the left-well population $P_l(t)$ for $N_{l2}=10$ and
$T_1=25$ ns (energy dissipation in the second qubit is neglected,
$T_1'=\infty$). The subset of the states used in this calculation is $141\le
n\le185$.
 Figure \ref{f12} shows the populations of the first 10 levels in
the left well; the populations of levels $k=10$ and 11 are not shown since
they are very close to zero.

\begin{figure}[tb]
\includegraphics[width=9cm]{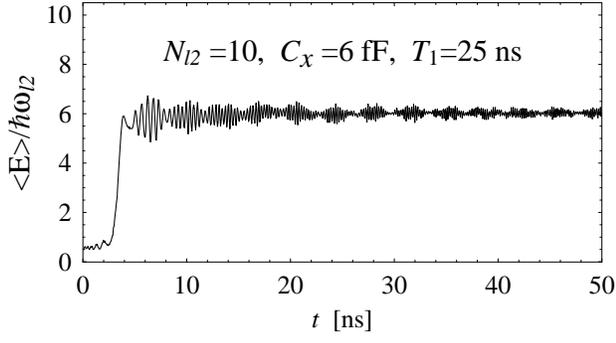}
\caption{Time dependence of the mean energy $\langle E\rangle$ of the second
qubit [Eq.\ (\protect\ref{3.28})] in units of $\hbar\omega_{l2}$ for
$N_{l2}=10$, $C_x=6$ fF, and $T_1=25$ ns. }
 \label{f13}\end{figure}

Figure \ref{f13} shows the time dependence of the qubit mean energy
 \be
\langle E(t)\rangle =\sum\nolimits_{n}E_nP_n(t)
 \e{3.28}
for the same parameters as in Figs.\ \ref{f11} and \ref{f12}. Comparing the
mean energy with the classical qubit energy for the same parameters (Fig.\
\ref{f7'}), we see that the two curves are similar, though classical energy
shows larger fluctuations.
 Note that the mean energy starts at $t=0$ from a
nonzero value equal to the qubit energy in the ground state
$\approx\hbar\omega_{l2}/2$.
 Even though the mean energy is significantly lower than the barrier height
(similar to the classical energy), the escape probability $P_s(t)=1-P_l(t)$
is nonzero in the quantum case (see Fig.\ \ref{f11}).

The time dependence of the switching probability (Fig.\ \ref{f11}) looks
quite irregular.
 This shows that the quantum behavior is rather complicated, similar to
the classical behavior discussed in Sec. \ref{IVB}.
 As seen in Figs.\ \ref{f12} and \ref{f13}, at $t\alt 3$ ns, when the
driving force is far from the resonance with the qubit, the population mainly
remains in the ground state.
 At this stage, there is no switching (see Fig.\ \ref{f11}).
Similar to the classical case, there is a fast qubit excitation (though
still almost without switching) between 3 ns and 4 ns (a little earlier than
the moment $t_c=4.8$ ns of classical resonance), while the main switching
happens much later, mostly at $t\simeq 16$ ns.

\begin{figure}[tb]
\includegraphics[width=8cm]{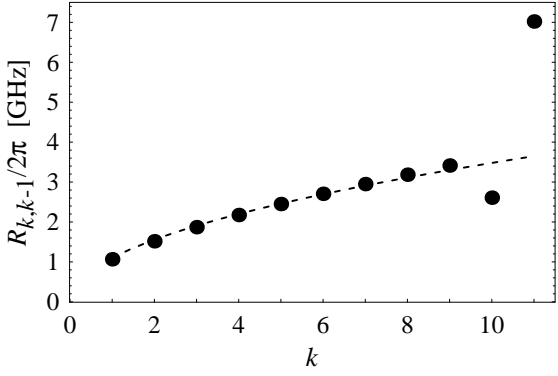}
\caption{Dots: Rabi frequencies $R_{k,k-1}/2\pi$ for the left-well
transitions at $t=t_c$, for $N_{l}=10$, $C_x=6$ fF, and $T_1=25$ ns. Dashed
line shows analytical dependence $1.1\sqrt{k}$ GHz.}
 \label{f13''}\end{figure}

 To understand the excitation mechanism, we show in Fig.\ \ref{f13''}
the Rabi frequencies $R_{k,k-1}=2 \zeta m''A|\delta_{k,k-1}|/\hbar$ for the
adjacent left-well transitions; $R_{k,k-1}$ is equal to the amplitude of
$V_{k,k-1}(t)/\hbar$ oscillations [actually, since $V(t)\propto
\ddot{\delta}_1(t)$ given by Eq.\ (\ref{3.23}) is significantly non-harmonic
in time, we need to use the amplitude of the resonant component].
    Since the amplitude $2A(t)$ of $\ddot{\delta}_1$ oscillations changes with time,
the Rabi frequencies also change with time. In Fig.\ \ref{f13''} we show the
values corresponding to the exact classical resonance,
$f_d(t_c)=\omega_{l2}/2\pi$, which happens at $t_c=4.8$ ns (then
$A=5.2\times 10^3$ ns$^{-2}$); for comparison, at $t\simeq 3$ ns the value
of $A$ and, correspondingly, the Rabi frequencies are approximately 10\%
smaller. (Notice that the moment $t_c$ scales with $T_1$, but the values of
Rabi frequencies at $t_c$ do not change with $T_1$.)
 For the levels not too close to the barrier top, one can use the
harmonic-oscillator relation \cite{lan77}
$|\delta_{k,k-1}|\approx\sqrt{k\hbar/2m''\omega_{l2}}\simeq 0.10 \sqrt{k}$
[for $N_{l2}=10$ and parameters of Eq.\ (\ref{2.16})], yielding
$R_{k,k-1}/2\pi\approx A(C_x/C)\sqrt{ 2km''/\hbar\omega_{l2}}/2\pi \simeq
1.1\sqrt{k}$ GHz. This formula fits well the numerical results in Fig.\
\ref{f13''} up to $k=9$; for higher levels anharmonicity becomes really
strong.

\begin{figure}[tb]
\includegraphics[width=8cm]{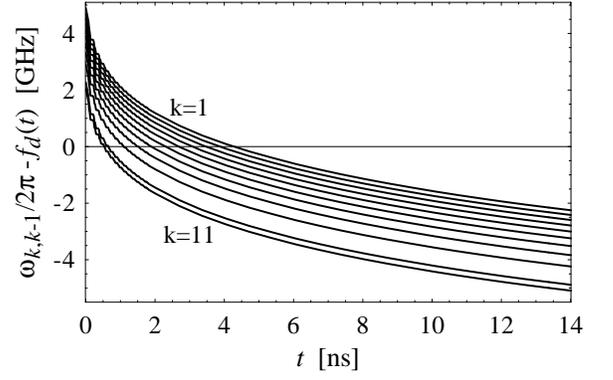}
\caption{Time dependence of the frequency detunings
$[\omega_{k,k-1}-\omega_d(t)]/2\pi$ of left-well transitions
$(k-1)\leftrightarrow k$ for $N_{l}=10$, $C_x=6$ fF, and $T_1=25$ ns.
 Curves from top to bottom correspond to $k=1,2\dots 11$.}
 \label{f13'}\end{figure}

 Figure \ref{f13'} shows the time dependence of the corresponding
detunings $\omega_{k,k-1}/2\pi -f_d(t)$.
 Though the exact resonance with the transition $0\leftrightarrow 1$
 ($\omega_{10}/2\pi=10.0$ GHz)
happens at $t=4.3$ ns, a significant excitation starts earlier, at
$t\approx3$ ns, when the detuning $\omega_{10}/2\pi -f_d(t)$ becomes less
than the Rabi frequency $R_{10} \simeq 1$ GHz.
 Since the Rabi frequency increases with $k$, while the detuning
first decreases and then increases with $k$ (after detuning changes sign),
the ground level population rapidly propagates to higher levels, until the
detuning becomes so large that the further excitation stops.
 As a result, at $t \agt 4$ ns the levels 0-2 become practically empty, while almost all the
population is transferred to levels 4-7. Similar to the classical case, the
excitation efficiency is significantly suppressed by the fact that driving
frequency $f_d(t)$ increases with time, while the level spacing decreases
with the level number $k$. Therefore, by the time at which levels $k\agt 5$
become populated, the further up-transitions are already out of resonance
(which happened for them earlier), becoming even farther off resonance with
increasing time.

 The escape (switching) in the quantum case can occur in several ways.
 The population which goes to the highest states $k=10$ and 11 is lost
rather fast (within a fraction of nanosecond), since those states interact
significantly with the right-well and delocalized states. However, in our
simulation transitions to these states are quite weak because of significant
detuning (even though the Rabi frequency is not much smaller than detuning,
the level occupations above $k=7$ decrease 4-10 times per level, so the
occupation of the level $k=10$ is already very small).

    Another switching mechanism is the following.
 With the increase of the driving frequency $f_d(t)$, it can
become resonant with transitions between non-adjacent states, thus
populating the states close or above the barrier, which cannot be populated
otherwise.
 In particular, more than half of the switching probability in
Fig.\ \ref{f11} is due to the sharp decrease of $P_l(t)$ between 16 ns and 18
ns, which happens because of the transition between the state $k=9\ (n=164)$
and the right-well state $n=168$ with the difference frequency of 13.2 GHz
(see the arrow in Fig.\ \ref{f10}).
 At $t=16$ ns the detuning for this transition is 0.7 GHz ($f_d=12.5$ GHz), which is
much smaller than other detunings between the levels of interest and is
comparable to the corresponding Rabi frequency $\simeq 0.5$ GHz.
 This value of the Rabi frequency is obtained as above,
taking into account the matrix element $|\delta_{164,168}|=0.048$ and the
value $A=5.8\times10^3$ ns$^{-2}$ at $t=16$ ns.

       Notice that in our numerical method the wavefunction is represented in
the basis of non-perturbed (time-independent) eigenstates. An alternative
way would be to diagonalize the Hamiltonian at each moment of time and use
the time-dependent eigenstates. Even though both methods are formally
equivalent, the second method would be more natural to use if the
dissipation is taken into account. In the time-dependent language an
important mechanism of escape is Landau-Zener tunneling through the barrier.
The perturbation $V(t)$ in the Hamiltonian (\ref{3.23}) is equivalent to
changing in time magnetic flux,
 \be
\phi_2\rightarrow\phi_2+\zeta\lambda m''\ddot{\delta}_1(t)/E_J,
 \e{3.29}
which changes $N_{l2}$ and leads to oscillations of the energy shift between
the comb of levels in the left well and the right-well comb. Because of
rather strong amplitude of these oscillations (in the example of Figs.\
\ref{f11}--\ref{f13'}, $N_{l2}$ oscillates between 9.15 to 11.2), each
left-well level crosses with several right-well levels during one
oscillation cycle.
 These (avoided) crossings lead to transitions between
the states in different wells (tunneling), the rate of which, according to
the Landau-Zener formula, \cite{lan77} is $\propto W^2$, where $W$ is the
minimal level splitting at the crossing.
 The values of $W$ increase exponentially with increase of $k$.
 Therefore, the Landau-Zener tunneling is a relatively slow (ineffective)
switching mechanism for all levels, except for the highest ones (such as
$k=10$ and 11). Besides the Landau-Zener mechanism, the escape from the left
well may also happen because the upper left-well states may become
delocalized (above the barrier) when the barrier $N_{l2}$ decreases in the
process of oscillations. We would also like to mention that the oscillations
of $\ddot\delta_1$ (and therefore of $N_{l2}$) are strongly non-harmonic. In
particular, the mentioned above range $9.15<N_{l2}<11.2$ remains constant
during long time interval $0< t < 0.4\, T_1$, because these extrema actually
do not correspond to the turning points of $\delta_1(t)$ oscillations;
instead, they correspond to the points of inflection $\delta_c'$ and
$\delta_c+2\pi$ of $U(x)$ [see Eq.\ (\ref{2.31})], as long as these points
are within the oscillation swing of $\delta_1(t)$.

We have performed extensive calculations of the switching probability
$P_s(t)$, varying the parameters $C_x$, $T_1$, and $N_{l2}$. We run
simulations within the time interval $[0,6\,T_1]$; after $6\,T_1$ the
first-qubit oscillations decay to a very low level, so the perturbation of
the second qubit is weak, and the further change of $P_s(t)$ is practically
negligible. Correspondingly, we define the total switching probability $P_s$
(which is the crosstalk error probability) as $P_s(6\, T_1)$.

Before discussing the results, let us briefly discuss our choice of the
restricted subset of $n_r$ states used for the numerical solution of Eq.\
(\ref{3.31}).
 First, we consider the eigenvalues of $\tilde{H}_{nn'}$ (\ref{3.30}) as
functions of $\ddot{\delta}_1$, and require that this dependence is
sufficiently close to the dependence obtained using the full Hamiltonian
(\ref{3.23}). Besides comparing the energies, we also compare the matrix
elements $\delta_{nn'}$ obtained using either the full or restricted space
(now only for the maximum and minimum values of $\ddot{\delta}_1$), and
require that the difference is below 1\%. The discussed below calculations
for the case $N_{l2}=5$ have been performed for the subset of $n_r=30$
states: $167\le n\le196$ (in this case there are 6 states localized in the
left well: $n=169,171,\dots,179$). For $N_{l2}=10$, as mentioned above, we
have used $n_r=45$ states: $141\le n\le 185$.
 We have also performed calculations with enlarged subsets (up to $n_r=150$)
and found that the switching probability varies irregularly with the number
of states in the subset; however, this variation is not significant, at
least in the case of low switching probability $P_s \alt 0.3$, which is the
range of our interest.

\begin{figure}[tb]
\includegraphics[width=8.3cm]{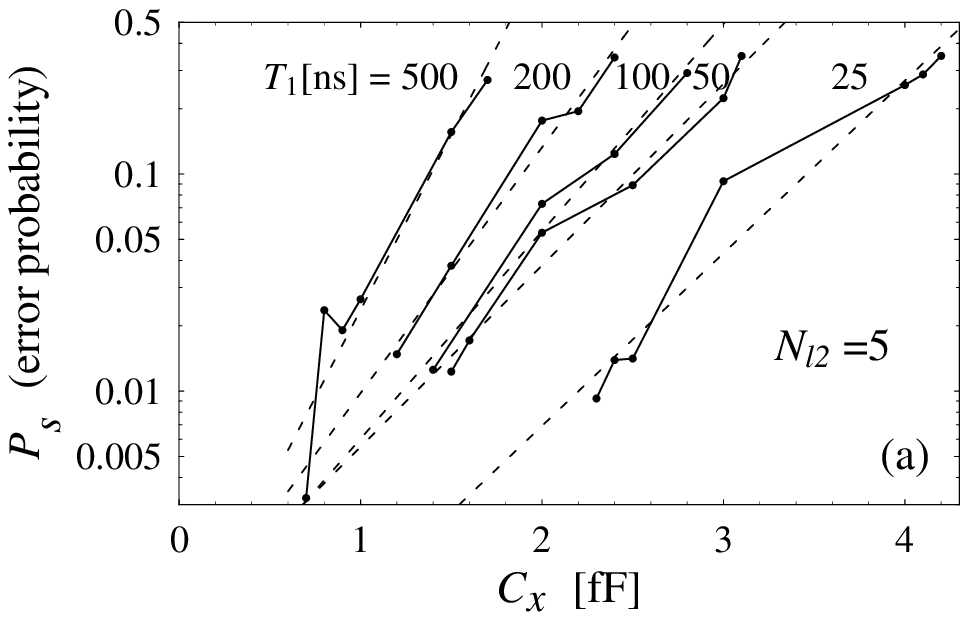}\\
\vspace{.2cm}
\includegraphics[width=8.3cm]{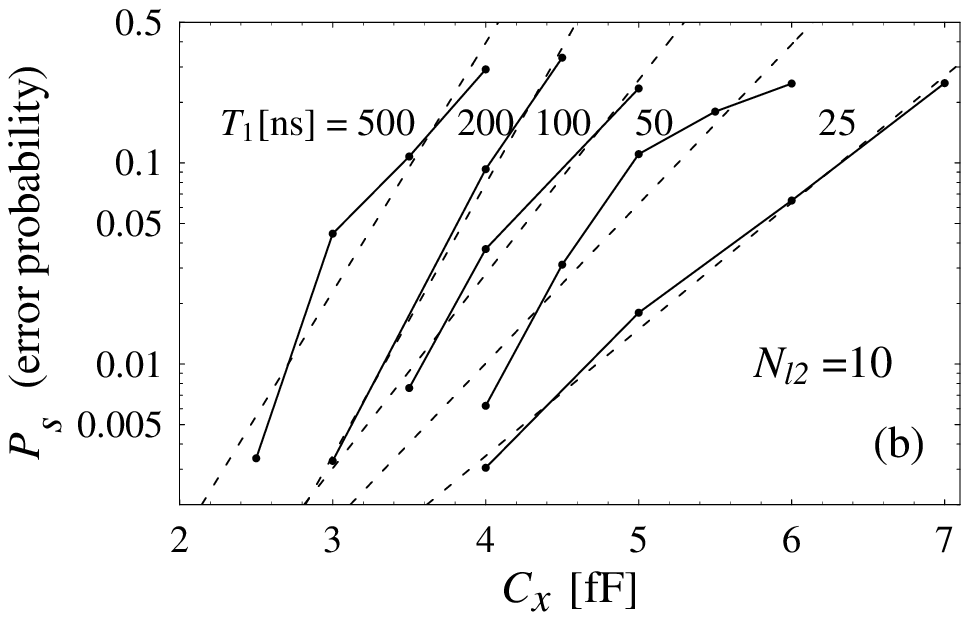}
\caption{The qubit switching (error) probability $P_s$ as a function of
coupling capacitance $C_x$ (in fF)  for $T_1=25$, 50, 100, 200, and 500 ns
for (a) $N_{l2}=5$  and (b) $N_{l2}=10$.
 The numerical data are represented by points, connected by solid
lines as guides for the eye. The dashed straight lines are results of the
least-squares fit (notice the logarithmic scale).}
 \label{f14}\end{figure}

 The results of numerical calculation of $P_s(C_x)$ dependence for several
values of $T_1$ and two values of $N_{l2}$ (5 and 10) are shown in Fig.\
\ref{f14}. Notice that the lines connecting the data points are not smooth
and sometimes are even nonmonotonous.
 This may be explained by complicated dynamics, similar to the
irregular behavior in the classical case. Despite the $P_s(C_x)$ dependence
in Fig.\ \ref{f14} is not smooth, we still see that the switching probability
$P_s$ decreases approximately exponentially with decrease of $C_x$.

\begin{figure}[tb]
\hspace{-0.2cm}\includegraphics[width=8.7cm]{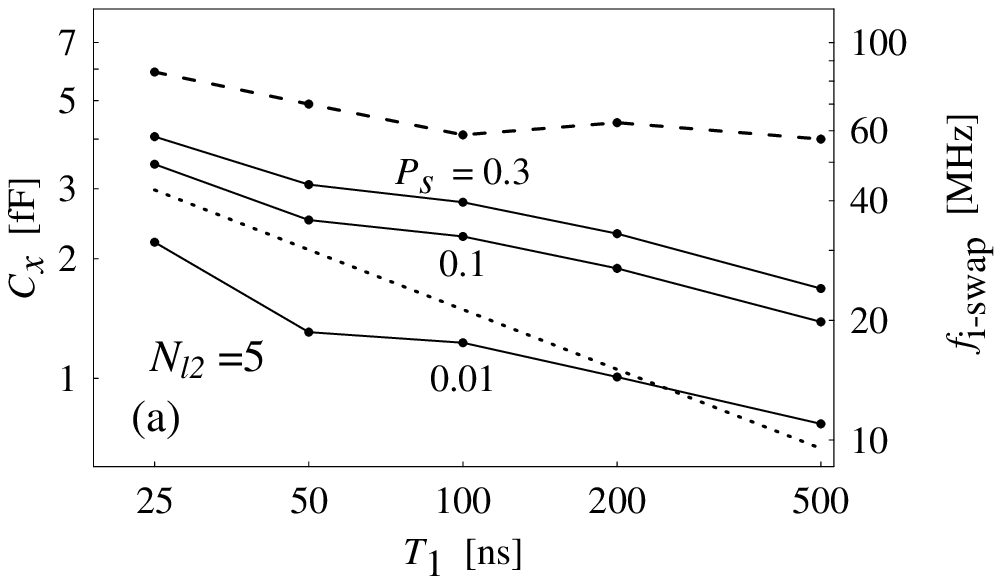}\\

\hspace{-0.2cm}\includegraphics[width=8.7cm]{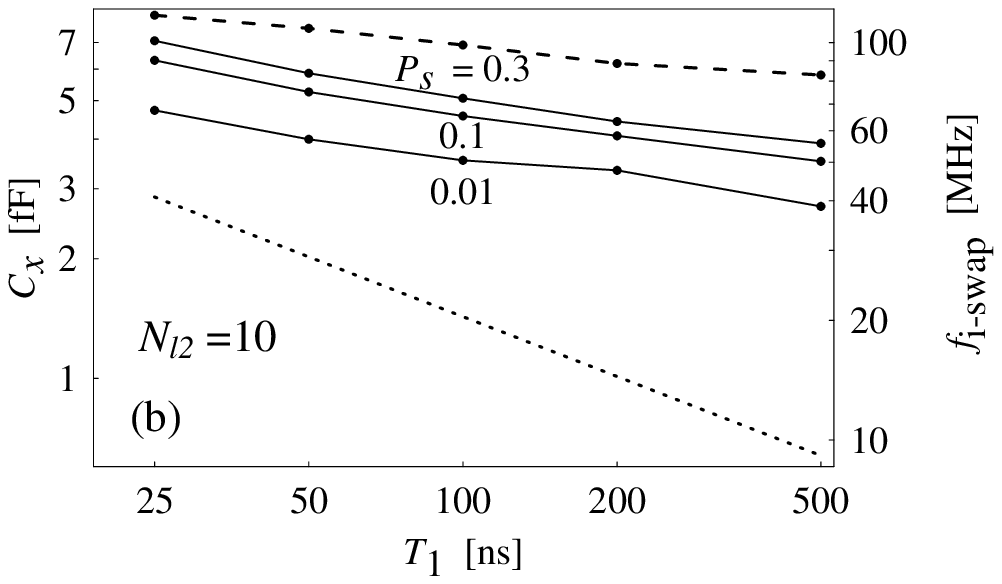}
 \caption{Solid
lines: log-log contour plots for the values of the error (switching)
probability $P_s=0.01$, 0.1, and 0.3 on the plane of relaxation time $T_1$
(in ns) and coupling capacitance $C_x$ (in fF) in the quantum model for (a)
$N_{l2}=5$ and (b) $N_{l2}=10$.
 The corresponding results for $C_{x,T}(T_1)$ in the classical models are
shown by the dashed lines (actual potential model) and the dotted
lines [oscillator model, Eq.\ (\ref{bound1})].
 The numerical data are represented by the points, connected by
lines as guides for the eye. The scale at the right corresponds to the
operation frequency of the two-qubit imaginary-swap \protect\cite{mcd05}
quantum gate.}
 \label{f16}\end{figure}

Using the linear (on the semilog scale) least-square fit for the results
shown in Fig.\ \ref{f14} (see dashed lines), we obtain the contour plots for
$P_s$ on the plane of $C_x$ and $T_1$ (see solid lines in Fig.\ \ref{f16}).
 The data in Fig.\ \ref{f16} can be fitted by straight
lines, yielding approximate power-law dependence  [(similar to Eq.\
(\ref{3.18})] for the threshold coupling capacitance $C_{x,T}$:
 \be
C_{x,T}(T_1)\approx B(P_s)T_1^{-\beta(P_s)},
 \e{3.32}
which now depends on the tolerable level $P_s$ of the measurement error
probability.
 The obtained numerical values of the parameters $B$ and $\beta$ in this formula
 are shown in Table
\ref{t1}.
 Notice that $\beta$ depends on $P_s$ quite weakly, but decreases
appreciably when $N_l$ changes from 5 to 10.
 The values of $\beta$ in Table \ref{t1} are greater than
the value $\beta=0.12$ obtained in the classical model with actual qubit
potential (Sec. \ref{IVB}), but less than the value $\beta=0.5$ in the
oscillator model [Eq.\ (\ref{bound1})]. This means that dependence on $T_1$
in Eq.\ (\ref{3.32}) in the quantum model is in-between those found in the
oscillator and actual-potential classical approaches.
 Figure \ref{f16} also shows a comparison between the results of classical
and quantum approaches. As one can see, both the oscillator (dotted line)
and actual-potential (dashed line) classical models give the limits for $C_x$
roughly similar to quantum results (within a factor of $\sim$ 2), with the
exception of the case of the oscillator model for $N_l=10$ and $T_1>100$ ns.
The value of the coupling capacitance $C_x$ determines the speed of two-qubit
quantum gates. The right scale of the vertical axis in Fig.\ \ref{f16}
converts $C_x$ into the operation frequency
$f_{\mbox{i-swap}}=(C_x/C)\,\omega_{10}/2\pi$ of the
imaginary-swap\cite{mcd05} gate.

 These results can be used for the design of phase-qubit-based
quantum gates.
 In particular, they give us the maximum allowed
coupling capacitance $C_{x}$ and hence the maximum gate operation speed, for
a particular tolerable value $P_s$ of the error due to crosstalk. An
important result of the quantum treatment is the exponential dependence of
the error probability on $C_x$ and a rather slow dependence on $T_1$. This
shows that the measurement crosstalk is not a big roadblock for the
fabrication of phase-qubit-based quantum gates with low decoherence and
sufficiently high operation speed.

\begin{table}[tb]
\begin{tabular}{c|ccc|ccc}
 $N_{l2}$&&5&&&10&\\
\hline
 $P_s$&0.01&0.1&0.3&0.01&0.1&0.3\\
\hline
 $B$&5.4&8.3&9.7&8.1&11&13\\
 $\beta$&0.32&0.29&0.28&0.18&0.19&0.20
 \end{tabular}
\caption{Parameters of Eq.\ (\protect\ref{3.32}) for the quantum model
($C_x$ is in fF and $T_1$ is in ns), limiting the coupling capacitance $C_x$
for several values of the error probability $P_s$ and dimensionless barrier
height $N_{l2}$.}
 \label{t1}\end{table}

The present quantum theory does not take into account dissipation in the
second qubit.
 The dissipation shortens the effective crosstalk time and thus
decreases the crosstalk error (the switching probability), similar to the
classical case discussed in Sec. \ref{IVB}.
 Thus, the present results give a lower bound for the maximum allowed $C_{x}$.
Taking into account the results of the classical model, one may expect
$\sim$ 30\% larger limit for the coupling capacitance (and two-qubit gate
frequency) for a quantum model with energy dissipation in the second qubit.

\section{Conclusions}
\label{VI}

The main goal of this paper has been to study the crosstalk between two
capacitively coupled flux-biased phase qubits after the measurement pulse.
 The first qubit, which escapes (switches) from the left to
the right well during the pulse, has been modeled classically. The first
qubit performs damped oscillations (with energy relaxation time $T_1$) with
increasing in time frequency $f_d(t)$; these oscillations perturb the
capacitively coupled second qubit.
 The dynamics of the second qubit (which is initially in the ground state) has been
treated both classically and quantum-mechanically.

 In the classical treatment of the second qubit, we have compared the
previously suggested \cite{mcd05} oscillator model, which allows for both
analytical and numerical analysis, with the model based on the exact
potential, which can be solved only numerically.
 Both models show a sharp resonant excitation of the second qubit.
 Though there is a certain similarity between the two models,
they significantly differ both quantitatively and qualitatively.
 In contrast to the oscillator model, the exact-potential model shows
nonlinear and irregular dynamics.
 The second qubit remains in the left well when the coupling
capacitance is sufficiently low, $C_x\le C_{x,T}$, but may escape (though
not certainly because of complicated dynamics) if $C_x>C_{x,T}$.
 We have obtained numerically the dependence $C_{x,T}(T_1)$ both in
absence and presence of dissipation in the second qubit for experimentally
relevant values of the barrier height.

For the quantum treatment we have developed an efficient numerical scheme,
which uses a subset of eigenstates of the unperturbed Hamiltonian.
 In this case, similarly to the classical case, a fast excitation of
the second qubit occurs at a moment when the driving frequency $f_d(t)$ is
somewhat below the transition frequency between the ground and first excited
states.
 However, in contrast to the classical case, the switching can now happen
even when the qubit mean energy is significantly lower than the barrier
height, either due to tunneling or due to excitation above the barrier.

The results for the switching (error) probability $P_s$ have been presented
as contour plots on the plane of coupling capacitance $C_x$ and relaxation
time $T_1$ (Fig.\ \ref{f16}).
 Such plots may be important for the design of quantum gates based on
phase qubits. Comparison of the results obtained in the quantum and
classical models shows that the classical models can be used for a crude
estimate of the crosstalk error; however, the difference becomes significant
for $T_1\agt 100$ ns.
 In the quantum approach the dissipation in the second qubit have been
neglected.
 However, by analogy with the classical case, one can expect that
the account of dissipation will not change the results significantly, though
it will somewhat increase the upper bound for the coupling capacitance,
above which the crosstalk error becomes intolerable. The model analyzed in
this paper assumes fixed coupling between the qubits. Future implementations
of experimental schemes with adjustable coupling (which can be practically
zeroed at the time of measurement) will significantly suppress the crosstalk
error mechanism and correspondingly allow for significant increase in the
operation frequency of the two-qubit quantum gates.

The present study can also be of relevance for problems in other fields
(e.g., laser chemistry), which consider excitation or escape from a
potential well by an oscillating driving force with changing in time
parameters.

    The work was supported by NSA and DTO under ARO grant W911NF-04-1-0204.


\begin{thebibliography}{99}
\bibitem{bla03-n}
For coupled current-biased phase qubits, this Hamiltonian was obtained,
e.g., in Refs.\ \protect\onlinecite{bla03,joh03,wu06}; similar Hamiltonian
for coupled charge qubits was obtained, e.g., in Ref.\
\protect\onlinecite{Pashkin}.
\bibitem{rounding} The chosen unrounded value for $N_{l1}$ corresponds to a rounded
value for another parameter [$(\phi_c-\phi_1)/\lambda$] which we used in the
code.
\bibitem{rem-gauge}
Alternatively, one can represent Eq.\ (\protect\ref{3.20}) as
$H(t)=H_0+V'(t)$, where $V'(t)$ is the term linear in $\hat{p}$.
 Note, however, that in this case (because of the ``vector potential''), the
generalized momentum $\hat{p}$ differs from the kinetic momentum
$m''\hat{\dot{\delta}}$ and hence $H_0$ [Eq.\ (\protect\ref{3.23})] does not
correspond to the classical energy [given by Eq.\ (\protect\ref{3.17})].
Thus, the gauge without ``vector potential'' (without terms linear in
$\hat{p}$) has the advantage that in contrast to other gauges, the
coefficients of $\psi_n(\delta)$ in the expansion (\protect\ref{3.24}) of the
wave function have the physical meaning of the probability amplitudes for the
energy eigenstates. \protect\cite{lam52}
\end{thebibliography}
\end{document}